\documentclass[12pt,journal,draftclsnofoot,oneside,onecolumn]{IEEEtran}

\usepackage[cmex10]{amsmath}
\usepackage{amssymb}
\usepackage{epsfig}
\usepackage{subfigure}
\usepackage{cite}
\usepackage{graphicx}
\usepackage{color}
\usepackage{bm}
\usepackage{booktabs}
\usepackage{gensymb}
\usepackage{mathrsfs}
\usepackage{xfrac}
\usepackage{algorithm}
\usepackage{algorithmic}
\newlength{\figwidth}
\newtheorem{lemma}{\it Lemma}

\newtheorem{corollary}{\it Corollary}
\newtheorem{remark}{\it Remark}
\newtheorem{proposition}{\it Proposition}
\makeatletter

\begin{document}

\title{STARS-ISAC: How Many Sensors Do We Need?}
\author{Zheng Zhang,~\IEEEmembership{Student Member,~IEEE}, Yuanwei Liu,~\IEEEmembership{Senior Member,~IEEE}, \\ Zhaolin Wang,~\IEEEmembership{Graduate Student Member,~IEEE},
Jian Chen,~\IEEEmembership{Member,~IEEE},
\thanks{Zheng Zhang and Jian Chen are with the School of Telecommunications Engineering, Xidian University, Xi'an 710071, China (e-mail: zzhang\_688@stu.xidian.edu.cn; jianchen@mail.xidian.edu.cn). Yuanwei Liu and Zhaolin Wang are with the School of Electronic Engineering and Computer Science,
Queen Mary University of London, London E1 4NS, U.K. (e-mail: yuanwei.liu@qmul.ac.uk; zhaolin.wang@qmul.ac.uk;).}
}
\maketitle

\begin{abstract}
  A simultaneously transmitting and reflecting surface (STARS) enabled integrated sensing and communications (ISAC) framework is proposed, where a novel bi-directional sensing-STARS architecture is devised to facilitate the \emph{full-space} communication and sensing. Based on the proposed framework, a joint optimization problem is formulated, where the Cram$\acute{\text{e}}$r-Rao bound (CRB) for estimating the 2-dimension direction-of-arrival of the sensing target is minimized. Two cases are considered for sensing performance enhancement. \textit{1) For the two-user case}, an alternating optimization algorithm is proposed. In particular, the maximum number of deployable sensors is obtained in the closed-form expressions. \textit{2) For the multi-user case}, an extended CRB (ECRB) metric is proposed to characterize the impact of the number of sensors on the sensing performance. Based on the proposed metric, a novel penalty-based double-loop (PDL) algorithm is proposed to solve the ECRB minimization problem. To tackle the coupling of the ECRB, a general decoupling approach is proposed to convert it to a tractable weighted linear summation form. Simulation results reveal that 1) the proposed PDL algorithm can achieve a near-optimal performance with consideration of sensor deployment; 2) without violating the communication under the quality of service requirements, reducing the receive antennas at the BS does not deteriorate the sensing performance; and 3) it is preferable to deploy more passive elements than sensors in terms of achieving optimal sensing performance.

\end{abstract}

\begin{IEEEkeywords}
  Beamforming design, integrated sensing and communications (ISAC), simultaneously transmitting and reflecting surface (STARS), sensor deployment.
\end{IEEEkeywords}

\section{Introduction}
With the commercialization of the fifth generation (5G) wireless networks, the 2030-oriented sixth generation (6G) wireless communication systems drew growing attention in both academia \cite{6G_academia} and industry \cite{6G_huawei,6G_samsung}. 6G seeks a fundamental paradigm shift in wireless network architecture, which breaks the physical boundaries of sensing and communications to support more emerging applications, such as extended reality (XR), auto-driving, and Metaverse. To realize this vision, a key enabling technique, integrated sensing and communication (ISAC), has been proposed to unify the two functions via the same hardware platform and signal processing module \cite{F.Liu_ISAC_JSAC,F.Liu_ISAC_TCOM}. To elaborate, through the dedicated co-designed framework, ISAC is capable of significantly enhancing the utilization efficiency of the network resources, thereby resulting in low implementation overheads. Furthermore, through deep integration, ISAC is also envisioned to realize mutual assistance and win-win benefit between the two functions \cite{Z.Zhang_VTM}.

To provide ubiquitous wireless connection with low energy consumption, reconfigurable intelligent surface (RIS) has emerged as another promising and cost-effective technique for future wireless networks \cite{M.Di.Renzo_RIS_JSAC,Yuanwei_RIS_magazine}. Technically, RIS can be regarded as a metasurface-based planar array, which is composed of lots of passive tunable elements. The electromagnetic response at each element can be proactively adjusted via an external smart controller, which aims to reconfigure the amplitude and phase shifts of the incident signal and thus realize a smart radio environment. However, since the conventional RIS can merely reflect the incident signals and provide half-space coverage, the design flexibility is stringently limited by its geographical location and panel orientation. As a remedy, a new RIS paradigm, namely simultaneously transmitting and reflecting surface (STARS), has been proposed \cite{Liu_magazine,X.Mu_STAR}. Compared to the conventional reflecting-only RIS, STARS can provide \emph{full-space} electromagnetic environment reconfiguration \cite{J.Xu_coupled_STAR}.


\subsection{Prior Works}
There have been lots of efforts devoted to the ISAC networks \cite{F.Liu_MIMO,F.Liu_Optimal_waveform,X.Liu_ISAC_beamforming,F.Liu_CRB,F.Liu_Resource}. More specifically, the authors of \cite{F.Liu_MIMO} devised a pair of antenna setups for multi-antenna ISAC systems, where a high-accuracy beampattern strategy is proposed to improve the sensing performance while guaranteeing the communication requirements. As a further advance, the authors of \cite{F.Liu_Optimal_waveform} proposed the optimal sensing waveform strategies, where the performance tradeoff between communication and sensing was investigated. To introduce more degrees-of-freedom (DoFs) for target sensing, a sophisticated ISAC framework was proposed in \cite{X.Liu_ISAC_beamforming}, where the independent radar waveforms and communication symbols were exploited to form the multiple beams for high-quality sensing. However, the aforementioned works only focused on the waveform design at the transmitter while neglecting the sensing performance imposed by the received echo at the receiver. To provide a comprehensive evaluation of the sensing performance, the authors of \cite{F.Liu_CRB} introduced the Cram$\acute{\text{e}}$r-Rao bound (CRB) as the sensing performance metric of an unbiased estimation at the receiver. Furthermore, the authors of \cite{F.Liu_Resource} developed a fairness-oriented unified resource allocation framework, where the BS was employed to carry out the device-free sensing services while satisfying communication QoS demands.

More recently, it is claimed that the proper exploitation of RISs in ISAC systems can further boost the sensing performance \cite{Y.Wang_ISAC_RIS,H.Luo_ISAC_RIS,H.Zhang_ISAC_RIS,X.Wang_ISAC_RIS,X.Song_IRS_ISAC}. In \cite{Y.Wang_ISAC_RIS}, the authors proposed a RIS-assisted ISAC framework, in which a RIS is employed to establish reliable line-of-sight (LoS) links for distance and velocity estimation. To support scenarios with multiple point-like targets, the authors of \cite{H.Luo_ISAC_RIS} developed a majorization-minimization algorithm for target tracking by collaboratively designing the transmit beampattern and RIS coefficients. In \cite{H.Zhang_ISAC_RIS}, the authors conceived a joint optimization scheme regarding the sensing waveform and the RIS coefficients from the perspective of sensing mutual information. Considering the practical restriction of discrete phase shifts, a constant-modulus sensing waveform was designed in \cite{X.Wang_ISAC_RIS}, in which the multi-user interference was minimized under the sensing CRB constraint. In \cite{X.Song_IRS_ISAC}, the authors considered utilizing the RIS to provide sensing services to the blind-zone target, where the CRB minimization problems were investigated in the cases of point target and extended target, respectively.

However, the direct combination of RIS and ISAC in the aforementioned works inevitably increased the number of reflections experienced by the echo signals, which restricted the sensing performance. To address this issue, the authors of \cite{X.Shao_IRS_ISAC} pioneered a RIS-self-sensing architecture, where the dedicated sensors are deployed at the RIS to carrying out the sensing functionality. Shortly thereafter, the authors of \cite{C.Zhong_ISAC} proposed a two-phase semi-passive RIS-assisted ISAC scheme, where the RIS supported the uplink communications in the first phase while carrying out the multi-user location sensing in the second one. Exploiting the same semi-passive sensing-at-RIS architecture, the authors of \cite{X.Hu_IRS_ISAC} studied the effect of sensing functionality on the communications, where the RIS was employed to sense the user location to facilitate the communication beamforming design. Most recently, there was a preliminary exploration of STARS-enabled ISAC networks in \cite{Z.Wang_STAR_ISAC}, where a sensing-at-STARS structure was proposed to achieve the 2-dimension direction-of-arrivals (DOAs) estimation.

\subsection{Motivations and Contributions}
Based on the aforementioned RIS-enabled ISAC works \cite{Y.Wang_ISAC_RIS,H.Luo_ISAC_RIS,H.Zhang_ISAC_RIS,X.Wang_ISAC_RIS,X.Song_IRS_ISAC,X.Shao_IRS_ISAC,C.Zhong_ISAC,X.Hu_IRS_ISAC,Z.Wang_STAR_ISAC}, we can obtain following two observations.
\begin{itemize}
  \item Although there have been a few works focusing on the RIS/STARS-enabled ISAC systems, the communication users and/or targets are only considered to be located on one side of the RIS/STARS\footnote{In the work \cite{Z.Wang_STAR_ISAC}, STARS is employed to divide the whole space into the communication region and sensing region, where the communication users and target are only situated on the corresponding half-space region.}. Whereas in the practical networks, the users and targets are probably in different geographical positions at different times, even on both sides of the RIS/STARS. Apparently, such a problem cannot be coped with by the existing schemes, which thus calls for a more general strategy for the RIS/STARS-enabled ISAC systems.
  \item For the sensor-at-RIS/STARS architecture \cite{X.Shao_IRS_ISAC,C.Zhong_ISAC,X.Hu_IRS_ISAC,Z.Wang_STAR_ISAC}, a critical endogenous problem has not been answered yet, i.e., should we deploy more sensors or passive elements (PEs) at the RIS/STARS? Given the fixed number of total elements of the RIS/STARS, deploying more sensors can increase the echo sampling resolution. Whereas deploying more PEs can introduce more spatial DoFs to favor both communication and sensing performance. Hence, there may exist a tradeoff between the number of sensors and PEs, which requires further investigation.
\end{itemize}

Motivated by the above observations, we propose a STARS-enabled ISAC framework, where the communication users and the targets are located on both sides of the STARS, with a particular focus on the tradeoff between the number of sensors and PEs. Our main contributions are summarized below.
\begin{itemize}
  \item We propose a novel bi-directional sensing-STARS architecture, where the micro-sized sensors with encapsulated antennas are integrated into the transparent substrate of STARS to provide full-space communication and sensing service. To tackle the energy/signal leakage issue in uplink STARS transmissions, a time switching (TS) protocol based two-phase scheme is proposed, where the STARS periodically switches between the reflection and transmission modes to support different users/targets. With this transmission framework, the closed-form CRB expression is derived as the sensing performance metric for estimating the 2-dimension DOAs.
  \item We first consider a two-user network. A CRB minimization problem is formulated subject to the communication quality of service (QoS) constraint of the ergodic achievable rate. To facilitate the optimization, the approximated ergodic rate is derived in the closed-form expression. Then, we propose an alternating optimization (AO) algorithm, where the optimal sensing waveform, transmit power, and reflection/transmission coefficients are alternately obtained by utilizing the standard semidefinite program (SDP) method. To provide further insights, the maximum number of sensors that can be deployed is derived in the closed-form expression, which unveils that the maximum number of sensors is only relevant to the QoS requirements of communications.
  \item We further consider a multi-user network, where a new metric of extended CRB (ECRB) is proposed to transform the impact of the number of sensors on the sensing accuracy into an explicit form. We aim to minimize the proposed ECRB under the communication QoS requirements. To efficiently solve the formulated mixed integer non-linear program (MINLP), a generic decomposing method is devised to transform the non-convex objective function of the ECRB into a weighted linear summation form of the constant ECRB matrices. Based on the transformation, a penalty-based double-loop (PDL) algorithm is proposed to solve the resultant non-convex optimization problem.
  \item Numerical results demonstrate the effectiveness and convergence of the proposed algorithms. It is also verified that the proposed PDL can enable a near-optimal allocation of the number of sensors. Besides, two insights are observed: 1) with the proposed bi-directional sensing-STARS architecture, reducing the number of receive antennas at the BS does not deteriorate the sensing performance provided that QoS requirements are satisfied; and 2) given a fixed total number of STARS elements, deploying more PEs at the STARS is more attractive than sensors in terms of sensing performance enhancement.
\end{itemize}

The organization of this paper is as follows. In Section \ref{Section_2}, we present the system model and performance metrics. In Section \ref{Section_3}, we conceive an AO algorithm to minimize CRB under the given number of sensors. Section \ref{Section_4} develops a PDL algorithm for the joint optimization of beamforming and the number of sensors. The numerical results are illustrated in Section \ref{Section_5}. Finally, the conclusion is presented in Section \ref{Section_6}

\textit{Notations:} we use the boldface capital $\mathbf{X}$ and lower-case letter $\mathbf{x}$ to represent matrix and vector, respectively. For any $N\times M$-dimensional matrix $\mathbf{X}\in\mathbb{C}^{N\times M}$, $\mathbf{X}^{T}$ and $\mathbf{X}^{H}$ denote the transpose and Hermitian conjugate operations. Similarly, $\text{rank}(\mathbf{X})$, $\text{Tr}(\mathbf{X})$, $\|\mathbf{X}\|$, $\|\mathbf{X}\|_{\text{F}}$ represent the rank value, trace value, spectral norm operation and Frobenius norm operation. $\mathbf{X}\succeq\mathbf{0}$ denotes that $\mathbf{X}$ is a positive semidefinite matrix, while $\mathbf{x}\sim \mathcal{CN}(0,\mathbf{X})$ denotes that $\mathbf{x}$ is a circularly symmetric complex Gaussian (CSCG) vector with zero mean and covariance matrix $\mathbf{X}$. For a matrix $\mathbf{X}$, $\text{vec}(\mathbf{X})$ and $\mathbf{X}^{-1}$ denote the vectorizing and inverse matrix operation. For any vector $\mathbf{x}$, $\text{diag}(\mathbf{x})$ denotes a diagonal matrix whose main diagonal elements equal to the elements of $\mathbf{x}$. $|x|$ and $\|\mathbf{x}\|$ denote the modulus of $x$ and the Euclidean norm of the vector $\mathbf{x}$, respectively. $\mathbb{E}($\textperiodcentered$)$ is the statistical expectation operation. $\mathbf{I}_{N}$ is a $N$-dimensional identity matrix, and $\mathbf{0}_{N}$ is a $N$-dimensional zero matrix. $\Re(\cdot)$ and $\Im(\cdot)$ denote the real component and imaginary component of the complex value. For any complex scalar $z$, $\tilde{z}$ denotes the conjugate of $z$. For any real scalar $x$,$\lfloor x\rfloor$ and $\lceil x\rceil$ denote the round-down and round-up operations. $\odot$ denotes the Hadamard product.

\section{System Model}\label{Section_2}

\begin{figure}[t]
  \centering
  \includegraphics[scale = 0.27]{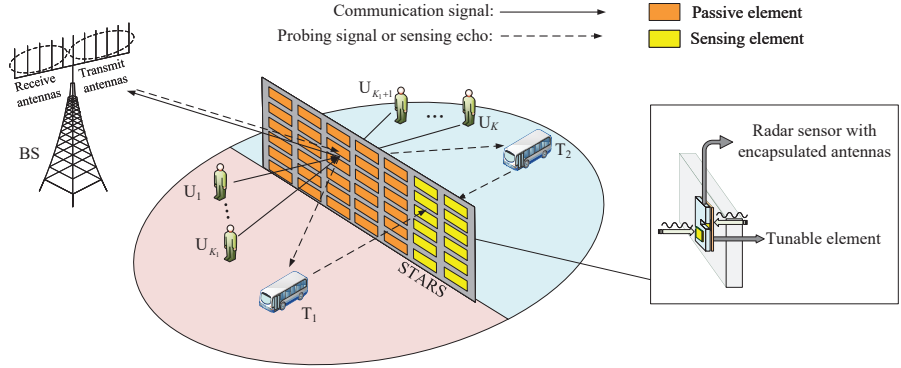}
  \caption{The bi-directional sensing-STARS enabled uplink ISAC network.}
  \label{Fig.1}
\end{figure}

\subsection{Network Description}
As shown in Fig. \ref{Fig.1}, we consider a STARS-enabled uplink ISAC network, where a STARS is deployed to establish reliable LoS links for $K$ blind-zone users $\{\text{U}_{1},\cdots,\text{U}_{K}\}$ to communicate with a BS while relaying the probing signal intended for the targets $\{\text{T}_{1},\text{T}_{2}\}$. The whole space is divided by the STARS into two separate region, each of which contains a target requiring estimation. It is assumed that the direct links of BS-users, users-targets, and BS-targets channels are blocked due to the obstacles. To mitigate the full-duplex self-interference at the BS, the BS is assumed to be equipped with an $M_{\text{t}}$-antenna transmit uniform linear array (ULA) and an $M_{\text{r}}$-antenna receive ULA \cite{F.Liu_CRB}, and all the users are single-antenna nodes. The STARS is composed of a uniform planar array (UPA) with $N=N_{\text{v}}N_{\text{h}}$ sub-wavelength elements, where $N_{\text{v}}$ and $N_{\text{h}}$ denote the number of elements located vertically and horizontally in the x-o-z plane, respectively.

To support the full-space communication and sensing, we propose a bi-directional sensing-STARS architecture, which divides the $N$ STARS elements into two parts. The former $N_{1}$ PEs are employed to support the uplink communication, and the latter $N_{2}=N-N_{1}$ elements (referred to as sensing elements) are equipped with the sensors for targets estimation. More specifically, the PEs operate in the reflection or transmission mode for information transmission \cite{Liu_magazine}. For each sensing element, a micro-sized low-cost sensor with antennas in packages is integrated inside the transparent substrate of STARS, where the adjacent tunable element operates in the full transmission mode, with the unit amplitude coefficient and zero phase-shift manipulation for the dual-sided incident signals \cite{Jiaqi.Xu_Dual-sided}. Thus, the sensor is cable of receiving full-space echo waves without suffering penetration attenuation caused by STARS element. All the inter-antenna/element distances are assumed to be sub-wavelength, so the adjacent channel reflected/transmitted by the STARS element can be deemed to be independent channels \cite{V.Arun_RFocus}.

The considered network is assumed to be a narrow-band system, where the BS and users transmit probing signal and communication signal in one coherence block of $T$ consecutive sample frames \cite{F.Liu_CRB}. All the channels are assumed to be static at one coherence block, but vary over different coherence blocks \cite{C.Zhong_ISAC}. The channel coefficients from PEs to the BS and $\text{U}_{k}$ are respectively denoted as $\mathbf{G}_{\text{r}}\in\mathbb{C}^{N_{1}\times M_{\text{r}}}$, $\mathbf{G}_{\text{t}}\in\mathbb{C}^{N_{1}\times M_{\text{t}}}$ and $\mathbf{h}_{k,\text{S}}\in\mathbb{C}^{N_{1}\times 1}$. All the channels are assumed to follow Rician fading model since STARS can be flexibly deployed to favor the LoS links. Hence, the communication channels $\mathbf{H}_{\text{c}}\in\{\mathbf{G}_{\text{r}},\mathbf{G}_{\text{t}},\mathbf{h}_{k,\text{S}}\}$ are modeled as
\begin{equation}
\label{1}
\mathbf{H}_{\text{c}}=L_{\text{c}}\left(\sqrt{\frac{\kappa }{1+\kappa}}\mathbf{\hat{H}}_{\text{c}}+\sqrt{\frac{1}{1+\kappa}}\mathbf{\tilde{H}}_{\text{c}}\right),
\end{equation}
where $\kappa$ denotes the Rician factor, $\mathbf{\hat{H}}_{\text{c}}$ represents the LoS component, and $\mathbf{\tilde{H}}_{\text{c}}$ denotes the non-LoS component. $L_{\text{c}}\!=\!\sqrt{\!L_{0}d_{\text{c}}^{-{\alpha_{\text{c}}}}}\!\in\!\{L_{\text{r}},L_{\text{t}},L_{k,\text{S}}\}$ denotes the corresponding path loss, where $d_{\text{c}}$ is the communication distance, $\alpha_{\text{c}}$ represents the path-loss exponent, and $L_{0}$ denotes the path loss at the reference distance of $1$ meter (m). For the sensing process, the probing signal and echo wave undergoes the PEs$\rightarrow$target$\rightarrow$sensors channels, which is modeled as
\begin{equation}
\label{2}
\mathbf{H}_{[\imath],\text{s}}=\alpha_{[\imath]}\mathbf{b}(\varphi_{[\imath]},\phi_{[\imath]})\mathbf{a}^{T}(\varphi_{[\imath]},\phi_{[\imath]}),\quad 1\leq \imath\leq 2,
\end{equation}
where $\alpha_{[\imath]}\in\mathbb{C}$ is the reflection coefficient containing the radar cross section (RCS) of $\text{T}_{\imath}$ and the round-trip path-loss, $\varphi_{[\imath]}$ denotes the azimuth angle of arrival/departure from the STARS to the target $\text{T}_{\imath}$, and $\phi_{[\imath]}$ denote the elevation angle of arrival/departure from the STARS to the target $\text{T}_{\imath}$. Note that $\mathbf{a}(\varphi_{[\imath]},\phi_{[\imath]})$ and $\mathbf{b}(\varphi_{[\imath]},\phi_{[\imath]})$ denote the steering vectors of PEs and sensors, where the $n$-th elements of $\mathbf{a}(\varphi_{[\imath]},\phi_{[\imath]})$ and $\mathbf{b}(\varphi_{[\imath]},\phi_{[\imath]})$ are given by \cite{C.Zhong_ISAC}
\begin{align}
\label{3}
\mathbf{a}(\varphi_{[\imath]},\phi_{[\imath]})[n]&=e^{\jmath\left[\bar{n}\pi\cos\phi_{[\imath]}\sin\varphi_{[\imath]}+(n-1-N_{\text{h}}\bar{n})\pi\sin\phi_{[\imath]}\right]},
\quad 1\leq n\leq N_{1},\\
\label{4}
\mathbf{b}(\varphi_{[\imath]},\phi_{[\imath]})[n]&=e^{\jmath\left[\bar{n}\pi\cos\phi_{[\imath]}\sin\varphi_{[\imath]}+(n-1-N_{\text{h}}\bar{n})\pi\sin\phi_{[\imath]}\right]},
\quad N_{1}+1\leq n\leq N,
\end{align}
where $\bar{n}=\lfloor\frac{n-1}{N_{\text{h}}}\rfloor$. To investigate the fundamental performance limit of the considered network, the perfect channel information state (CSI) is assumed for all the channels.



\subsection{Signal Model}




Without loss of generality, we focus on the transmission at the $t$-th time frame and propose a TS-based framework, which equally divides the each time frame into following two phases.

\subsubsection{Phase I} The PEs operates in the reflection mode and users $\mathcal{K}_{1}\triangleq\{\text{U}_{1},\cdots,\text{U}_{K_{1}}\}$ transmit the communication signal $c_{[1]}(t)=\sum_{k\in\mathcal{K}_{1}}\sqrt{P_{k}}c_{k}(t)$ with $\mathbb{E}\{|c_{k}(t)|^{2}\}=1$ to the PEs. Meanwhile, the BS exploits the multiple beams to send the dedicated probing signal $\mathbf{s}_{[1]}(t)=\sum_{j=1}^{I_{[1]}}\check{\mathbf{s}}_{j}(t)\in\mathbb{C}^{M\times 1}$ with a general-rank covariance matrix $\mathbf{R}_{[1],\text{s}}=\mathbb{E}\{\mathbf{s}_{[1]}(t)\mathbf{s}_{[1]}^{H}(t)\}$. On receiving the omnidirectional signal, the PEs reflects the communication signal to the BS while combining the $c_{[1]}(t)$ and $\mathbf{s}_{[1]}(t)$ as a new probing signal to perform estimation.

\subsubsection{Phase II} The PEs operates in the transmission mode. The users $\mathcal{K}_{2}\triangleq\{\text{U}_{K_{1}+1},\cdots,\text{U}_{K}\}$ send the communication signal $c_{[2]}(t)=\sum_{k\in\mathcal{K}_{2}}\sqrt{P_{k}}c_{k}(t)$ to the PEs. Meanwhile, the BS transmits probing signal $\mathbf{s}_{[2]}(t)=\sum_{j=1}^{I_{[2]}}\check{\mathbf{s}}_{j}(t)\in\mathbb{C}^{M\times 1}$ ($\mathbf{R}_{[2],\text{s}}=\mathbb{E}\{\mathbf{s}_{[2]}(t)\mathbf{s}_{[2]}^{H}(t)\}$) to the PEs. At the STARS, the PEs is exploited to transmit the $c_{[2]}(t)$ to the BS while reconfiguring the probing signal to detect $\text{T}_{2}$.

Accordingly, the received signal at the BS in the $\imath$-th phase is given by
\begin{equation}
\label{5}
\mathbf{y}_{[\imath]}(t)=\mathbf{h}_{k,\text{S}}^{H}\bm{\Theta}_{\text{r}/\text{t}}\mathbf{G}_{\text{r}}\sum_{k\in\mathcal{K}_{\imath}}\sqrt{P_{k}}c_{k}(t)+\mathbf{n}_{[\imath]}(t),
\end{equation}
where $\mathbf{n}_{[\imath]}(t)\sim\mathcal{CN}(0,\sigma^{2}\mathbf{I}_{M_{\text{r}}})$ denotes the $\imath$-th phase additive white Gaussian noise (AWGN) at the BS, and $\bm{\Theta}_{\text{r}/\text{t}}=\bm{\Theta}_{\text{r}}$ in case of $\imath=1$ while denoting $\bm{\Theta}_{\text{t}}$ in case of $\imath=2$. Thereinto, the reflection/transmission coefficient matrix is defined as $\bm{\Theta}_{\text{r}/\text{t}}=\text{diag}(\mathbf{u}_{\text{r}/\text{t}})$ with $\mathbf{u}_{\text{r}/\text{t}}=[\sqrt{\beta_{\text{r}/\text{t},1}}e^{\jmath\theta_{\text{r}/\text{t},1}},\dots, \sqrt{\beta_{\text{r}/\text{t},N}}e^{\jmath\theta_{\text{r}/\text{t},N}}]^{T}$. To recover $c_{k}(t)$, we assume that a unit-norm linear combination vector $\mathbf{v}_{[\imath],k}\in\mathbb{C}^{M_{\text{r}}\times 1}$ is employed at the BS. The extracted signal for $\text{U}_{k}$ is given by
\begin{equation}
\label{6}
\mathbf{y}_{[\imath]}(t)\mathbf{v}_{[\imath],k}=\underbrace{\mathbf{h}_{k,\text{S}}^{H}\bm{\Theta}_{\text{r}/\text{t}}\mathbf{G}_{\text{r}}\mathbf{v}_{[\imath],k}\sqrt{P_{k}}c_{k}(t)}_{\text{desired signal}}+
\underbrace{\mathbf{h}_{j,\text{S}}^{H}\bm{\Theta}_{\text{r}/\text{t}}\mathbf{G}_{\text{r}}\mathbf{v}_{[\imath],k}\sum\limits_{j\neq k, j\in \mathcal{K}_{\imath}}\sqrt{P_{j}}c_{j}(t)}_{\text{interference}}+
\mathbf{n}_{[\imath]}(t)\mathbf{v}_{[\imath],k}.
\end{equation}
While for the sensing, the equivalent probing signal from PEs at the $t$-th time frame of the $\imath$-th phase is given by
\begin{align}
\label{7}
\mathbf{x}_{[\imath]}(t)=
\begin{cases}
\mathbf{H}_{\text{U},\text{S}}\sqrt{\mathbf{\bar{P}}}\mathbf{c}(t)+\mathbf{G}_{\text{t}}\mathbf{s}_{[1]}(t), \quad &\text{if} \ \imath=1,\\
\mathbf{G}_{\text{t}}\mathbf{s}_{[2]}(t), \quad &\text{if} \ \imath=2,\\
\end{cases}
\end{align}
where $\mathbf{H}_{\text{U},\text{S}}=[\mathbf{h}_{1,\text{S}},\cdots,\mathbf{h}_{K_{1},\text{S}}]$, $\mathbf{\bar{P}}=\text{diag}[P_{1},\cdots,P_{K_{1}}]$, and $\mathbf{c}(t)=[c_{1},\cdots,c_{K_{1}}]^{T}$. Accordingly, the covariance matrix of $\mathbf{x}_{[\imath]}(t)$ is given by
\begin{align}
\label{8}
\mathbf{R}_{[\imath],\mathbf{x}}=\mathbb{E}[\mathbf{x}_{[\imath]}(t)\mathbf{x}_{[\imath]}(t)^{H}]=
\begin{cases}
\mathbf{H}_{\text{U},\text{S}}\mathbf{\bar{P}}\mathbf{H}_{\text{U},\text{S}}^{H}+
\mathbf{G}_{\text{t}}\mathbf{R}_{[1],\text{s}}\mathbf{G}_{\text{t}}^{H}, \quad &\text{if} \ \imath=1,\\
\mathbf{G}_{\text{t}}\mathbf{R}_{[2],\text{s}}\mathbf{G}_{\text{t}}^{H}, \quad &\text{if} \ \imath=2,\\
\end{cases}
\end{align}
Thus, the received echo wave at the sensors over $T$ consecutive time frames of the $\imath$-th phase is given by
\begin{equation}
\label{9}
\mathbf{Y}_{[\imath],\text{s}}=\mathbf{H}_{[\imath],\text{s}}\bm{\Theta}_{\text{r}/\text{t}}\mathbf{X}_{[\imath]}+\mathbf{N}_{[\imath]},
\end{equation}
where $\mathbf{X}_{[\imath]}=[\mathbf{x}_{[\imath]}(1),\cdots,\mathbf{x}_{[\imath]}(T)]$ and $\mathbf{N}_{[\imath]}=[\mathbf{n}_{[\imath]}(1),\cdots,\mathbf{n}_{[\imath]}(T)]$.

\subsection{Performance Metric}
As stated above, the achievable rate at the BS in the $\imath$-th phase to decode $c_{k}(t)$ is expressed as
\begin{equation}
\label{10}
R_{[\imath],k}=\frac{1}{2}\log_{2}\left(1+\frac{P_{k}|\mathbf{h}_{k,\text{S}}^{H}\bm{\Theta}_{\text{r}/\text{t}}\mathbf{G}_{\text{r}}\mathbf{v}_{[\imath],k}|^2}
{\sum\limits_{j\neq k,j\in \mathcal{K}_{\imath}}P_{j}
|\mathbf{h}_{j,\text{S}}^{H}\bm{\Theta}_{\text{r}/\text{t}}\mathbf{G}_{\text{r}}\mathbf{v}_{[\imath],k}|^2+\sigma^2}\right),\quad 1\leq \imath\leq 2,
\end{equation}

For the sensing, we focus on the CRB performance with unknown parameters $\bm{\varsigma}_{[\imath]}=[\varphi_{[\imath]},\phi_{[\imath]},\\ \Re(\alpha_{[\imath]}),\Im(\alpha_{[\imath]})]\in\mathbb{R}^{4\times 1}$. To facilitate deriving CRB expression, we vectorize equation \eqref{9}, which can be rewritten as
\begin{equation}
\label{11}
\mathbf{y}_{[\imath],\text{s}}=\text{vec}(\mathbf{Y}_{[\imath],\text{s}})=
\text{vec}\left(\mathbf{H}_{[\imath],\text{s}}\bm{\Theta}_{\text{r}/\text{t}}\mathbf{X}_{[\imath]}\right)+\mathbf{n}_{[\imath]},
\end{equation}
where $\mathbf{n}_{[\imath]}=\text{vec}(\mathbf{N}_{[\imath]})\sim\mathcal{CN}(0,\sigma^{2}\mathbf{I}_{MT})$. Let $\mathbf{q}_{[\imath]}=\text{vec}\left(\mathbf{H}_{[\imath],\text{s}}\bm{\Theta}_{\text{r}/\text{t}}\mathbf{X}_{[\imath]}\right)$, the Fisher information matrix (FIM) $\mathbf{F}_{[\imath]}\in\mathbb{R}^{4\times 4}$ for estimating $\bm{\varsigma}_{[\imath]}$ can be expressed as a Jacobian matrix with the $h$-th row and $v$-th column element being given by \cite{FIM}
\begin{align}
\label{12}\nonumber
\mathbf{F}_{[\imath]}[h,v]&=2\Re\left(\frac{\partial\mathbf{q}_{[\imath]}^{H}}{\partial\bm{\varsigma}_{[\imath],h}}\mathbf{R}_{\mathbf{n}_{[\imath]}}^{-1}
\frac{\partial\mathbf{q}_{[\imath]}}{\partial\bm{\varsigma}_{[\imath],v}}\right)+
\text{Tr}\left(\mathbf{R}_{\mathbf{n}_{[\imath]}}^{-1}\frac{\partial\mathbf{R}_{\mathbf{n}_{[\imath]}}}{\partial\bm{\varsigma}_{[\imath],h}}
\mathbf{R}_{\mathbf{n}_{[\imath]}}^{-1}\frac{\partial\mathbf{R}_{\mathbf{n}_{[\imath]}}}{\partial\bm{\varsigma}_{[\imath],v}}\right)\\
&=\frac{2}{\sigma^{2}}\Re\left(\frac{\partial\mathbf{q}_{[\imath]}^{H}}{\partial\bm{\varsigma}_{[\imath],h}}
\frac{\partial\mathbf{q}_{[\imath]}}{\partial\bm{\varsigma}_{[\imath],v}}\right), \quad 1\leq h,v \leq 4,
\end{align}
where $\mathbf{R}_{\mathbf{n}_{[\imath]}}=\sigma^{2}\mathbf{I}_{MT}$. Accordingly, we can repartition $\mathbf{F}_{[\imath]}$ as
\begin{equation}
\label{13}
\mathbf{F}_{[\imath]}=\begin{bmatrix}\mathbf{J}_{\bm{\Psi}_{[\imath]}\bm{\Psi}_{[\imath]}} & \mathbf{J}_{\bm{\Psi}_{[\imath]}\bm{\alpha}_{[\imath]}}\\
\mathbf{J}_{\bm{\Psi}_{[\imath]}\bm{\alpha}_{[\imath]}}^{T} & \mathbf{J}_{\bm{\alpha}_{[\imath]}\bm{\alpha}_{[\imath]}} \end{bmatrix},
\end{equation}
where $\bm{\Psi}_{[\imath]}=[\varphi_{[\imath]},\phi_{[\imath]}]$, $\bm{\alpha}_{[\imath]}=[\Re(\alpha_{\imath}),\Im(\alpha_{\imath})]$, while the specific expressions of $\mathbf{J}_{\bm{\Psi}_{[\imath]}\bm{\Psi}_{[\imath]}}$, $\mathbf{J}_{\bm{\Psi}_{[\imath]}\bm{\alpha}_{[\imath]}}$ and $\mathbf{J}_{\bm{\alpha}_{[\imath]}\bm{\alpha}_{[\imath]}}$ are given in Appendix A. Then, with the inverse formula of the second order matrix, we can derive the CRB expression of the $\imath$-th phase with regard to $\bm{\Psi}_{[\imath]}$ \cite{CRB_derive} as
\begin{equation}
\label{14}
\text{CRB}(\bm{\Psi}_{[\imath]})=\left[\mathbf{J}_{\bm{\Psi}_{[\imath]}\bm{\Psi}_{[\imath]}}-\mathbf{J}_{\bm{\Psi}_{[\imath]}\bm{\alpha}_{[\imath]}}
\mathbf{J}_{\bm{\alpha}_{[\imath]}\bm{\alpha}_{[\imath]}}^{-1}\mathbf{J}_{\bm{\Psi}_{[\imath]}\bm{\alpha}_{[\imath]}}^{T}\right]^{-1}.
\end{equation}

\section{Beamforming Optimization Under Fixed Sensor Number}\label{Section_3}

In this section, we concentrate on joint sensing waveform and communication beamforming optimization with the fixed number of sensors. To draw instructive insights for practical system design, we consider a special network setup of two users, and utilize the ergodic rate as the average communication performance metric. The the closed-form approximation expression of ergodic achievable rate is derived. Accordingly, an AO algorithm to efficiently solve the non-convex problem. 

\subsection{Problem Formulation}
Since inter-user interference is non-existent in the two-user case, the achievable rate at the BS in the $\imath$-th phase to decode $c_{k}(t)$ is rewritten to
\begin{equation}
\label{15}
R_{[\imath],k}=\frac{1}{2}\log_{2}\left(1+\frac{P_{k}|\mathbf{h}_{k,\text{S}}^{H}\bm{\Theta}_{\text{r}/\text{t}}\mathbf{G}_{\text{r}}\mathbf{v}_{[\imath],k}|^2}
{\sigma^2}\right).
\end{equation}

%

To provide the generalised insights to the CRB optimization, we employ the ergodic achievable rate as the average performance metric for the communication. Accordingly, we target at minimizing the CRB performance for DOA $\bm{\Psi}_{[\imath]}$ at each phase. Subject to the average QoS requirements of users, the joint optimization of the transmit power at the users, reflection/transmission coefficients of the PEs, the receive beamforming and the sensing waveform at the BS is considered. The optimization problem in the $\imath$-th phase is formulated as follows.
\begin{subequations}
\begin{align}
\label{16a} &\min\limits_{\mathbf{P}_{[\imath]},\bm{\Theta}_{\text{r}/\text{t}},\mathbf{R}_{[\imath],\text{s}},\mathbf{v}_{[\imath],k}}\quad \text{Tr}(\text{CRB}(\bm{\Psi}_{[\imath]})) \\
\label{16b}&\quad\text{s.t.} \quad \text{Tr}(\mathbf{R}_{[\imath],\text{s}})\leq P_{\text{BS},\text{max}},\\
\label{16c}&\quad\quad\quad\,\,  P_{k} \leq P_{\text{U},\text{max}},\quad 1\leq k\leq K, \\
\label{16d}&\quad\quad\quad\,\,  \mathbb{E}\{R_{k}\}\geq R_{\text{er},\text{t}},\\
\label{16e}&\quad\quad\quad\,\, \|\mathbf{v}_{[\imath],k}\|^{2}= 1,\quad 1\leq k\leq K, \\
\label{16f}&\quad\quad\quad\,\, \theta_{\text{r},n}, \theta_{\text{t},n}\in[0,2\pi], \ 1\leq n\leq N,\\
\label{16g}&\quad\quad\quad\,\, \beta_{\text{r},n}, \beta_{\text{t},n}\in[0,1], \  1\leq n \leq N,
\end{align}
\end{subequations}
where $\mathbf{P}_{[1]}=[P_{1},\cdots,P_{K_{1}}]$, $\mathbf{P}_{[2]}=[P_{K_{1}+1},\cdots,P_{K}]$, $R_{\text{er},\text{t}}$ denotes the ergodic QoS rate of users, $P_{\text{U},\text{max}}$ denotes the maximal transmit power at the users, and $P_{\text{BS},\text{max}}$ denotes the transmit power budget at the BS. \eqref{16b} and \eqref{16c} denote the transmit power constraint at the BS and users' sides, respectively; \eqref{16d} represents the ergodic QoS constraint of users; \eqref{16e} denotes the normalization constraint of the receive beamforming; \eqref{16f} and \eqref{16g} are the phase-shift and amplitude constraints of the PEs. Notably, the intractable expression of CRB and the non-convex constraints \eqref{16d} and \eqref{16e} make problem (16) non-convex and challenging to solve. In the following, we consider optimizing the sensing waveform and communication beamforming in an alternating manner.

\subsection{Problem Reformulation}
Before handling the challenging problem, we can observe that $\mathbf{v}_{[\imath],k}$ only exists in the constraint \eqref{16d} and has no direct influence on the CRB performance, which indicates that only the feasible $\mathbf{v}_{[\imath],k}$ are required. For maximum compliance with QoS constraint, the optimal receive beamforming is given by $\mathbf{v}_{[\imath],k}=\frac{(\mathbf{h}_{k,\text{S}}^{H}\bm{\Theta}_{\text{r}/\text{t}}\mathbf{G}_{\text{r}})^{H}}
{\|\mathbf{h}_{k,\text{S}}^{H}\bm{\Theta}_{\text{r}/\text{t}}\mathbf{G}_{\text{r}}\|}$, to obtain the best communication performance. Hence, \eqref{15} can be transformed into
$R_{[\imath],k}=\frac{1}{2}\log_{2}\left(1+\frac{P_{k}\|\mathbf{h}_{k,\text{S}}^{H}\bm{\Theta}_{\text{r}/\text{t}}\mathbf{G}_{\text{r}}\|^2}
{\sigma^2}\right)$. With this in mind, we further rewrite $\|\mathbf{\hat{h}}_{k,\text{S}}^{H}\bm{\Theta}_{\text{r}/\text{t}}\mathbf{\hat{G}}_{\text{r}}\|^{2}$ as $\text{Tr}(\mathbf{\hat{H}}_{k,\text{S}}\mathbf{\hat{H}}_{k,\text{S}}^{H}\mathbf{U}_{\text{r}/\text{t}})$, with definition of $\mathbf{U}_{\text{r}/\text{t}}=\mathbf{u}_{\text{r}/\text{t}}\mathbf{u}_{\text{r}/\text{t}}^{H}$ and $\mathbf{\hat{H}}_{k,\text{S}}=\text{diag}(\mathbf{\hat{h}}_{k,\text{S}}^{H})\mathbf{\hat{G}}_{\text{r}}$. Then, we resort Lemma \ref{Lemma_1} to obtain the approximation expression of $\mathbb{E}\{R_{k}\}$.
\begin{lemma}\label{Lemma_1}
    The ergodic achievable rate in \eqref{16d} can be approximated as
    \begin{align}
    \nonumber
    \mathbb{E}\{R_{[\imath],k}\}&\approx\frac{1}{2}\log_{2}\!\!\bigg(\!1\!+\!\frac{P_{k}L_{k,\text{S}}^{2}L_{\text{r}}^{2}}{\sigma^2(1\!+\!\kappa)^{2}}
    \bigg[\kappa^{2}\text{Tr}(\mathbf{\hat{H}}_{k,\text{S}}\mathbf{\hat{H}}_{k,\text{S}}^{H}\mathbf{U}_{\text{r}/\text{t}})\!+\!
    \kappa \text{Tr}(\mathbf{\hat{G}}_{\text{r}}\mathbf{\hat{G}}_{\text{r}}^{H}\mathbf{U}_{\text{r}/\text{t}})+\\ \label{17}
    \kappa &M_{\text{r}}\text{Tr}\left(\text{diag}(\mathbf{\hat{h}}_{k,\text{S}}^{H})\mathbf{U}_{\text{r}/\text{t}}\text{diag}(\mathbf{\hat{h}}_{k,\text{S}})\right)
    +M_{\text{r}}\text{Tr}(\mathbf{U}_{\text{r}/\text{t}})\bigg]\bigg).
    \end{align}
\end{lemma}
\begin{IEEEproof}
See Appendix B.
\end{IEEEproof}

The approximation accuracy can be verified in Fig. \ref{Fig.3}. To proceed, we further introduce Lemma \ref{Lemma_2} to convexify the non-convex objective function \eqref{16a}.
\begin{lemma}\label{Lemma_2}
    According to \cite{S.Boyd_convex}, we have that for any $\mathbf{X}\succeq\mathbf{0}$ and $\mathbf{Y}\succeq\mathbf{0}$, if $\mathbf{X}\succeq\mathbf{Y}$ is guaranteed, the inequality of $\text{Tr}(\mathbf{X}^{-1})\leq\text{Tr}(\mathbf{Y}^{-1})$ holds.
\end{lemma}

As such, with the fact that FIM matrix $\left[\mathbf{J}_{\bm{\Psi}_{[\imath]}\bm{\Psi}_{[\imath]}}-\mathbf{J}_{\bm{\Psi}_{[\imath]}\bm{\alpha}_{[\imath]}}
\mathbf{J}_{\bm{\alpha}_{[\imath]}\bm{\alpha}_{[\imath]}}^{-1}\mathbf{J}_{\bm{\Psi}_{[\imath]}\bm{\alpha}_{[\imath]}}^{T}\right]$ is positive semidefinite, we can introduce an auxiliary variable $\mathbf{E}_{[\imath]}\succeq\mathbf{0}$ to equivalently transform \eqref{16a} into $\text{Tr}(\mathbf{E}_{[\imath]}^{-1})$, with satisfying following linear matrix inequality (LMI) constraint.
\begin{equation}
    \label{18}
    \begin{bmatrix}\mathbf{J}_{\bm{\Psi}_{[\imath]}\bm{\Psi}_{[\imath]}}-\mathbf{E}_{[\imath]} & \mathbf{J}_{\bm{\Psi}_{[\imath]}\bm{\alpha}_{[\imath]}}\\
\mathbf{J}_{\bm{\Psi}_{[\imath]}\bm{\alpha}_{[\imath]}}^{T} & \mathbf{J}_{\bm{\alpha}_{[\imath]}\bm{\alpha}_{[\imath]}} \end{bmatrix}
\succeq \mathbf{0}.
\end{equation}
With the transformations above, problem (16) can be reformulated as follows.
\begin{subequations}
\begin{align}
\label{19a} &\min\limits_{\mathbf{E}_{[\imath]},\mathbf{U}_{\text{r}/\text{t}},\mathbf{R}_{[\imath],\text{s}},\mathbf{P}_{[\imath]}}\quad \text{Tr}(\mathbf{E}_{[\imath]}^{-1}) \\
\label{19b}&\quad\text{s.t.} \quad \eqref{16b},\eqref{16c},\eqref{18},\\
\label{19c}\nonumber&\quad\quad\quad \frac{P_{k}L_{k,\text{S}}^{2}L_{\text{r}}^{2}}{(1+\kappa)^{2}}\bigg[\kappa^{2}\text{Tr}(\mathbf{\hat{H}}_{k,\text{S}}\mathbf{\hat{H}}_{k,\text{S}}^{H}\mathbf{U}_{\text{r}/\text{t}})\!+\!
    \kappa M_{\text{r}}\text{Tr}\left(\text{diag}(\mathbf{\hat{h}}_{k,\text{S}}^{H})\mathbf{U}_{\text{r}/\text{t}}\text{diag}(\mathbf{\hat{h}}_{k,\text{S}})\right)\\
    &\quad\quad\quad +\kappa \text{Tr}(\mathbf{\hat{G}}_{\text{r}}\mathbf{\hat{G}}_{\text{r}}^{H}\mathbf{U}_{\text{r}/\text{t}})
    +M_{\text{r}}\text{Tr}(\mathbf{U}_{\text{r}/\text{t}})\bigg] \geq \sigma^2(2^{2R_{\text{er},\text{t}}}-1),\\
\label{19d}&\quad\quad\quad \mathbf{E}_{[\imath]}\succeq\mathbf{0},\quad \mathbf{U}_{\text{r}/\text{t}} \succeq \mathbf{0},\\
\label{19e}&\quad\quad\quad  \mathbf{U}_{\text{r}/\text{t}}[n,n] \leq 1, \quad 1\leq n\leq N,\\
\label{19f}&\quad\quad\quad  \text{rank}(\mathbf{U}_{\text{r}/\text{t}})=1.
\end{align}
\end{subequations}

\subsection{Joint Beamforming Optimization}

\subsubsection{Transmit power and sensing waveform optimization} With the fixed $\mathbf{U}_{\text{r}/\text{t}}$, the problem (19) is reduced to the following subproblem.
\begin{subequations}
\begin{align}
\label{20a} &\min\limits_{\mathbf{R}_{[\imath],\text{s}},\mathbf{E}_{[\imath]},\mathbf{P}_{[\imath]}}\quad \text{Tr}(\mathbf{E}_{[\imath]}^{-1}) \\
\label{20b}&\quad\text{s.t.} \quad \eqref{16b},\eqref{16c},\eqref{18},\eqref{19c}, \\
\label{20c}&\quad\quad\quad \mathbf{E}_{[\imath]}\succeq\mathbf{0},
\end{align}
\end{subequations}
which is a SDP and can be optimally solved by the convex toolbox, e.g., CVX.

\subsubsection{Reflection/Transmission coefficients optimization} With fixed $\{\mathbf{R}_{[\imath],\text{s}},\mathbf{P}_{[\imath]}\}$, problem (19) is rewritten as
\begin{subequations}
\begin{align}
\label{21a} &\min\limits_{\mathbf{E}_{[\imath]},\mathbf{U}_{\text{r}/\text{t}}}\quad \text{Tr}(\mathbf{E}_{[\imath]}^{-1}) \\
\label{21b}&\quad\text{s.t.} \quad \eqref{18},\eqref{19c}-\eqref{19f}.
\end{align}
\end{subequations}
To handle the non-convex LMI constraint \eqref{18}, we adopt the singular value decomposition (SVD) to equivalently convert the quadratic terms $\{\mathbf{J}_{\bm{\Psi}_{[\imath]}\bm{\Psi}_{[\imath]}},\mathbf{J}_{\bm{\Psi}_{[\imath]}\bm{\alpha}_{[\imath]}},\mathbf{J}_{\bm{\alpha}_{[\imath]}\bm{\alpha}_{[\imath]}}\}$ to the tractable forms. Specifically, by decomposing $\mathbf{R}_{\mathbf{x}_{[\imath]}}$ into $\sum_{q}\mathbf{s}_{q}\mathbf{d}_{q}$, we have
\begin{align}\nonumber
\bm{\Theta}_{\text{r}/\text{t}}\mathbf{R}_{\mathbf{x}_{[\imath]}}\bm{\Theta}_{\text{r}/\text{t}}^{H}&=
\sum_{q}\text{diag}(\mathbf{s}_{q})\mathbf{u}_{\text{r}/\text{t}}\mathbf{u}_{\text{r}/\text{t}}^{H}\text{diag}(\mathbf{d}_{q})\\  \label{22}
&=\sum_{q}\mathbf{S}_{q}\mathbf{u}_{\text{r}/\text{t}}\mathbf{u}_{\text{r}/\text{t}}^{H}\mathbf{D}_{q}=\sum_{q}\mathbf{S}_{q}\mathbf{U}_{\text{r}/\text{t}}\mathbf{D}_{q}.
\end{align}
Then, we substitute \eqref{22} into FIM matrix, the constraint \eqref{18} becomes convex with respect to $\mathbf{U}_{\text{r}/\text{t}}$. While for the non-convex constraint \eqref{19f}, we employ the penalty-based rank-one relaxation approach \cite{X.Yu_rank-one}, which exploits the successive convex approximation (SCA) technique to relax the equivalent rank-one constraint $\text{Tr}(\mathbf{U}_{\text{r}/\text{t}})-\|\mathbf{U}_{\text{r}/\text{t}}\|_{2}=0$ as a convex penalty term in objective function \eqref{21a}. Accordingly, the problem (21) is reformulated as
\begin{subequations}
\begin{align}
\label{23a} &\min\limits_{\mathbf{E}_{[\imath]},\mathbf{U}_{\text{r}/\text{t}}}\quad \text{Tr}(\mathbf{E}_{[\imath]}^{-1})-\frac{1}{2\rho_{1}}\left[\text{Tr}(\mathbf{U}_{\text{r}/\text{t}})-
\|\mathbf{U}_{\text{r}/\text{t}}^{[n-1]}\|_{2}-\text{Tr}(\mathbf{\bar{u}}_{\text{r}/\text{t}}^{[n-1]}
(\mathbf{\bar{u}}_{\text{r}/\text{t}}^{[n-1]})^{H}(\mathbf{U}_{\text{r}/\text{t}}-\mathbf{U}_{\text{r}/\text{t}}^{[n-1]}))\right], \\
\label{23b}&\quad\text{s.t.} \quad \eqref{18},\eqref{19c}-\eqref{19e},
\end{align}
\end{subequations}
where $\mathbf{U}_{\text{r}/\text{t}}^{[n-1]}$ denotes the optimized result in the $(n-1)$-th iteration, $\mathbf{\bar{u}}_{\text{r}/\text{t}}^{[n-1]}$ denotes the leading eigenvector of $\mathbf{U}_{\text{r}/\text{t}}^{[n-1]}$, and $\rho_{1}$ represents the penalty factor. The resultant problem (23) is a convex program, where the rank-one solution can be optimally obtained when $p$ is sufficiently small \cite[Proposition 2]{X.Yu_rank-one}. The specific SCA algorithm to optimize $\mathbf{U}_{\text{r}/\text{t}}$ is summarized in \textbf{Algorithm \ref{SCA-rank-one}}.

\begin{algorithm}[t]
    \caption{SCA algorithm for rank-one solution.}
    \label{SCA-rank-one}
    \begin{algorithmic}[1]
        \STATE{Initialize initial $\mathbf{U}_{\text{r}/\text{t}}^{[n-1]}$ and $p_{1}$ with $n=1$. Set a convergence accuracy $\epsilon_{1}$ and calculate the leading eigenvector $\mathbf{\bar{u}}_{\text{r}/\text{t}}^{[n-1]}$.}
        \REPEAT
        \STATE{ update $\mathbf{U}_{\text{r}/\text{t}}^{[n]}$ by solving problem (23). }
        \STATE{ update the leading eigenvector $\mathbf{\bar{u}}_{\text{r}/\text{t}}^{[n]}$.}
        \STATE{set $n=n+1$ and $\rho_{1}=\frac{\rho_{1}}{c_{1}}$ ($c_{1}>1$).}
        \UNTIL{ the penalty term in objective function \eqref{23a} drops below $\epsilon_{1}$.}
    \end{algorithmic}
\end{algorithm}

\begin{remark} (\textit{MLE Validation})
    In this paper, we consider employing the maximum likelihood estimation (MLE) approach in \cite[Appendix E]{X.Song_IRS_ISAC} to obtain the estimated DoA $\bm{\Psi}_{[\imath]}^{\text{es}}$ under the optimized waveform, where the the correctness of the proposed CRB optimization framework is demonstrated in Fig. \ref{Fig.5}.
\end{remark}

\begin{corollary} (\textit{Maximal Number of Sensor Deployment}) \label{Corollary_1}
    For the special case of single receive antenna, i.e., $M_{\text{r}}=1$, the optimal reflection/transmission coefficients and the maximal number of sensors to be deployed can be derived as following closed-form expressions.
    \begin{align}
    &\begin{cases}
    \theta_{\text{r},n}^{\text{opt}} = \angle\mathbf{\hat{h}}_{1,\text{S}}[n]-\angle\mathbf{\hat{g}}_{\text{r}}[n], \quad \beta_{\text{r},n}^{\text{opt}} = 1 \\ \label{24}
    \theta_{\text{t},n}^{\text{opt}} = \angle\mathbf{\hat{h}}_{2,\text{S}}[n]-\angle\mathbf{\hat{g}}_{\text{r}}[n], \quad \beta_{\text{t},n}^{\text{opt}} = 1,
    \end{cases}\\ \label{25}
    &N_{2}^{\text{max}} = N- \left\lceil\frac{1}{2\kappa^{2}}\Bigg[\sqrt{(2\kappa+1)^{2}+
    4\kappa^{2}\frac{\sigma^{2}(2^{2R_{\text{er},\text{t}}}-1)(1+\kappa)^2}{P_{\text{U},\text{max}}
    L_{k,\text{S}}^{2}L_{\text{r}}^{2}}}-
    (2\kappa+1)\Bigg]\right\rceil,
    \end{align}
    where $\mathbf{\hat{g}}_{\text{r}}$ is the degenerated channel of $\mathbf{\hat{G}}_{\text{r}}$.
\end{corollary}
\begin{IEEEproof}
See Appendix C.
\end{IEEEproof}

\subsection{Overall Algorithm}
\begin{algorithm}[t]
    \caption{AO algorithm.}
    \label{AO}
    \begin{algorithmic}[1]
        \STATE{Initialize initial $\mathbf{U}_{\text{r}/\text{t}}^{[l-1]}$ and $\text{Tr}(\text{CRB}(\bm{\Psi}_{[\imath]}))^{[l-1]}$ with $l=1$. Set a convergence accuracy $\epsilon_{2}$.}
        \REPEAT
        \STATE{ update the optimal receive beamforming $\mathbf{v}_{[\imath],k}^{\text{opt}}=\frac{(\mathbf{h}_{k,\text{S}}^{H}\bm{\Theta}_{\text{r}/\text{t}}\mathbf{G}_{\text{r}})^{H}}
{\|\mathbf{h}_{k,\text{S}}^{H}\bm{\Theta}_{\text{r}/\text{t}}\mathbf{G}_{\text{r}}\|}$.}
        \STATE{ update optimal $\{\mathbf{R}_{[\imath],\text{s}}^{[l]},\mathbf{P}_{[\imath]}^{[l]}\}$ by solving problem (20).}
        \STATE{ update optimal $\mathbf{U}_{\text{r}/\text{t}}^{[l]}$ by carrying out \textbf{Algorithm \ref{SCA-rank-one}}.}
        \STATE{ set $l = l+1$ and calculate $\text{Tr}(\text{CRB}(\bm{\Psi}_{[\imath]}))^{[l]}$.}
        \UNTIL{$|\text{Tr}(\text{CRB}(\bm{\Psi}_{[\imath]}))^{[l]}-\text{Tr}(\text{CRB}(\bm{\Psi}_{[\imath]}))^{[l-1]}|\leq \epsilon_{2}$.}
    \end{algorithmic}
\end{algorithm}

The overall algorithm is summarized in \textbf{Algorithm \ref{AO}}, which optimizes the sensing waveform and reflection/transmission coefficients alternatively. By denoting the CRB value at $l$-th iteration as a function of $\mathbf{R}_{[\imath],\text{s}}$ and $\mathbf{U}_{\text{r}/\text{t}}$, i.e., $g(\mathbf{R}_{[\imath],\text{s}},\mathbf{P}_{[\imath]},\mathbf{U}_{\text{r}/\text{t}})$, the following inequality always holds
\begin{align}
\label{26} g(\mathbf{R}_{[\imath],\text{s}}^{[l-1]},\mathbf{P}_{[\imath]}^{[l-1]},\mathbf{U}_{\text{r}/\text{t}}^{[l-1]}) \overset{(a)}{\geq} g(\mathbf{R}_{[\imath],\text{s}}^{[l]},\mathbf{P}_{[\imath]}^{[l]},\mathbf{U}_{\text{r}/\text{t}}^{[l-1]})\overset{(b)}{\geq}   g(\mathbf{R}_{[\imath],\text{s}}^{[l]},\mathbf{P}_{[\imath]}^{[l]},\mathbf{U}_{\text{r}/\text{t}}^{[l]}),
\end{align}
where inequality signs $(a)$ and $(b)$ hold because the optimal sensing waveform and optimal reflection/transmission coefficients are both guaranteed in the step $4$ and step $5$ at the same AO iteration. Meanwhile, since $g(\mathbf{R}_{[\imath],\text{s}},\mathbf{P}_{[\imath]},\mathbf{U}_{\text{r}/\text{t}})$ is continuous over the compact feasible set of problem, there exists a finite positive number that serves as a lower bound on the objective value. This proves that our proposed AO algorithm remains non-increasing over the iterations. On the other hand, the computational complexity of AO algorithm mainly relies on solving SDP problems (20) and (23). The overall complexity based on the interior-point method is given by $\mathcal{O}\left(\log(\frac{1}{\epsilon_{2}})\left((M_{\text{t}}^{2}+5)^{3.5}+\log(\frac{1}{\epsilon_{1}})(N_{1}^{2}+4)^{3.5}\right)\right)$ \cite{Z.Luo_complexity}.

\section{How Many Sensors Do We Need?}\label{Section_4}
In this section, we consider the general multi-user system with the joint optimization of beamforming design and the number of sensors. By modifying the traditional CRB expression, a new metric of ECRB is proposed, which can evaluate the sensing performance while taking the sensors' deployment into the consideration. Based on the proposed ECRB, a PDL algorithm is devised to jointly optimize the ISAC waveform, reflection/transmission coefficients and the number of PEs/sensors.

\subsection{Extended CRB Derivation}
To facilitate the optimization of sensor number, we define two $N$-dimensional matrices $\mathbf{A} = \text{diag}[\mathbf{I}_{N_{1}},\mathbf{0}_{N_{2}}]$ and $\mathbf{B} = \text{diag}[\mathbf{0}_{N_{1}},\mathbf{I}_{N_{2}}]$, where $\mathbf{A}$ and $\mathbf{B}$ are the element selection matrices for PEs and sensors, respectively. With the definition above, we can rewrite the steering vector $\mathbf{a}(\varphi_{[\imath]},\phi_{[\imath]})$ and $\mathbf{b}(\varphi_{[\imath]},\phi_{[\imath]})$ as the extended form.
\begin{align}
\label{27} \mathbf{a}(\varphi_{[\imath]},\phi_{[\imath]}) = \mathbf{A}\bm{\varepsilon}(\varphi_{[\imath]},\phi_{[\imath]})  \quad \mathbf{b}(\varphi_{[\imath]},\phi_{[\imath]}) = \mathbf{B}\bm{\varepsilon}(\varphi_{[\imath]},\phi_{[\imath]}).
\end{align}
Here, $\bm{\varepsilon}(\varphi_{[\imath]},\phi_{[\imath]})\in\mathbb{C}^{N\times 1}$ denotes the steering vector of the STARS, $n$-th element of which is defined in \eqref{3} with $1\leq n\leq N$. For convenience of denotation, we abbreviate $\bm{\varepsilon}(\varphi_{[\imath]},\phi_{[\imath]})$ as $\bm{\varepsilon}_{[\imath]}$ in the following. Then, we introduce Proposition \ref{Proposition_1} to derive the expression of the extended FIM matrix.
\begin{proposition}\label{Proposition_1}
    The $h$-th row and $v$-th column element of extended FIM matrix is given by
    \begin{align}
    \label{28}
    \mathbf{F}_{[\imath]}[h,v]=\frac{2TC_{[\imath]}^{(h,v)}}{\sigma^{2}}\text{Tr}\left(\mathbf{B}\mathbf{\bar{C}}_{\bm{\varsigma}_{[\imath]}[v]}\mathbf{A}\bm{\Theta}^{\text{r}/\text{t}}\mathbf{R}_{\mathbf{X}_{[\imath]}}
        \bm{\Theta}_{\text{r}/\text{t}}^{H}\mathbf{A}\mathbf{\bar{C}}_{\bm{\varsigma}_{[\imath]}[h]}^{H}\mathbf{B}\right), \, 1\leq h,v\leq 4,
    \end{align}
    where
    \begin{align}
    \label{29}
    C_{[\imath]}^{(i,j)}=\begin{cases}
    |\alpha_{[\imath]}|^{2}, \!\! &\text{if} \ 1\leq i,j\leq 2,\\
    \tilde{\alpha}_{[\imath]}, \!\! &\text{if} \ \overline{\iota}=3, \underline{\iota}\leq2,\\
    \jmath\tilde{\alpha}_{[\imath]}, \!\! &\text{if} \ \overline{\iota}=4, \underline{\iota}\leq2,\\
    1, \!\! &\text{if} \ 3\leq i,j\leq 4,
    \end{cases}
    \quad \mathbf{\bar{C}}_{\bm{\varsigma}_{[\imath]}[i]}=\begin{cases}
    \frac{\partial\bm{\varepsilon}_{[\imath]}}{\partial \bm{\Psi}_{[\imath]}[i]}\bm{\varepsilon}_{[\imath]}^{T}+\bm{\varepsilon}_{[\imath]}
\frac{\partial\bm{\varepsilon}_{[\imath]}^{T}}{\partial \bm{\Psi}_{[\imath]}[i]} \!\! &\text{if} \ 1\leq i\leq 2,\\
    \bm{\varepsilon}_{[\imath]}\bm{\varepsilon}_{[\imath]}^{T}, \!\! &\text{if} \ 3\leq i\leq 4,
    \end{cases}
    \end{align}
    where $\overline{\iota}=\max\{i,j\}$ and $\underline{\iota}=\min\{i,j\}$, and $\bm{\Theta}_{\text{r}/\text{t}}\in\mathbb{C}^{N\times N}$ is the reflection/transmission coefficient matrix of STARS.
\end{proposition}
\begin{IEEEproof}
See Appendix D.
\end{IEEEproof}
Therefore, the ECRB expression can be derived by substituting the extended FIM matrix expression into \eqref{14}. Similarly, the extended achievable rate expression can be expressed as  $R_{[\imath],k}=\frac{1}{2}\log_{2}\left(1+\frac{P_{k}|\mathbf{h}_{k,\text{S}}^{H}\bm{\Theta}_{\text{r}/\text{t}}\mathbf{A}\mathbf{G}_{\text{r}}\mathbf{v}_{[\imath],k}|^2}
{\sum\nolimits_{j\neq k,j\in\mathcal{K}_{\imath}}P_{j}
|\mathbf{h}_{k,\text{S}}^{H}\bm{\Theta}_{\text{r}/\text{t}}\mathbf{A}\mathbf{G}_{\text{r}}\mathbf{v}_{[\imath],k}|^2+\sigma^2}\right)$ with $\mathbf{h}_{k,\text{S}}\in\mathbb{C}^{N\times 1}$ and $\mathbf{G}_{\text{r}}\in\mathbb{C}^{N\times M_{\text{r}}}$.


\subsection{Problem Formulation}
With the derivations above, we aim to minimize the ECRB value for estimating $\bm{\Psi}_{[\imath]}$ of each phase under the QoS constraints, by jointly optimizing the transmit power at the users, the reflection/transmission coefficients of the PEs, the number of PEs/sensors at the STARS, the receive beamforming and sensing waveform at the BS. Based on the definitions of $\mathbf{U}_{\text{r}/\text{t}}=\mathbf{u}_{\text{r}/\text{t}}\mathbf{u}_{\text{r}/\text{t}}^{H}$ and $\mathbf{H}_{k,\text{S}}=\text{diag}(\mathbf{h}_{k,\text{S}}^{H})\mathbf{G}_{\text{r}}$, the problem formulation in the $\imath$-th phase is given by
\begin{subequations}
\begin{align}
\label{30a} \min_{\mathbf{E}_{[\imath]},\mathbf{P}_{[\imath]},\mathbf{U}_{\text{r}/\text{t}},
\atop
\mathbf{V}_{[\imath],k},\mathbf{A},\mathbf{B},\mathbf{R}_{[\imath],\text{s}}} &\text{Tr}(\mathbf{E}_{[\imath]}^{-1}) \\
\label{30b}\text{s.t.}\quad &\eqref{16b},\eqref{16c},\eqref{18},\eqref{19d}-\eqref{19f},\\
\label{30c}&P_{k}\text{Tr}(\mathbf{A}\mathbf{H}_{k,\text{S}}\mathbf{V}_{[\imath],k}\mathbf{H}_{k,\text{S}}^{H}\mathbf{A}\mathbf{U}_{\text{r}/\text{t}})\!\geq\!
\gamma_{\text{t}}\Big(\!\!\sum\limits_{j\neq k,j\in\mathcal{K}_{\imath}}\!\!\!\!\!\!P_{j}
\text{Tr}(\mathbf{A}\mathbf{H}_{j,\text{S}}\mathbf{V}_{[\imath],k}\mathbf{H}_{j,\text{S}}^{H}\mathbf{A}\mathbf{U}_{\text{r}/\text{t}})\!+\!\sigma^2\Big),\\
\label{30d}&\mathbf{A}[n,n]\in\{0,1\}\ ,\mathbf{B}[n,n] \in\{0,1\}, \quad 1\leq n\leq N, \\
\label{30e}&\mathbf{A}+\mathbf{B}=\mathbf{I}_{N},\\
\label{30f}&\text{Tr}(\mathbf{V}_{[\imath],k})=1, \quad \mathbf{V}_{[\imath],k}\succeq \mathbf{0},
\end{align}
\end{subequations}
where $\gamma_{\text{t}}=(2^{2R_{\text{t}}}-1)$ with $R_{\text{t}}$ representing the QoS requirements of users, and $\mathbf{V}_{[\imath],k}=\mathbf{v}_{[\imath],k}(\mathbf{v}_{[\imath],k})^{H}$. \eqref{30d} and \eqref{30e} denote the integer variable constraints for selection matrices. Note the problem (30) is a MINLP that cannot be optimally solved by conventional convex optimization methods, except for exhaustive search. To strike a balance between optimality and complexity, we propose a PDL algorithm obtain the near-optimal solution of problem (30), which optimizes the constructed augmented Lagrangian (AL) problem in the inner loop while updating the penalty factor in the outer loop.


\subsection{Augmented Lagrangian Problem Construction}
To convert problem (30) to a tractable form, we introduce the auxiliary variables $[p_{1},\cdots,p_{N-1}]$, which satisfies $p_{n}\in\{0,1\}$ and $\sum_{n=1}^{N-1}p_{n}=1$. Then, we can equivalently rewrite \eqref{28} as
\begin{align}
\label{31} \mathbf{F}_{[\imath]}[h,v]=\frac{2TC_{[\imath]}^{(h,v)}}{\sigma^{2}}\sum_{n=1}^{N-1}p_{n}\text{Tr}\left(\mathbf{\bar{F}}_{n,v}\bm{\Theta}^{\text{r}/\text{t}}\mathbf{R}_{\mathbf{X}_{[\imath]}}
        \bm{\Theta}_{\text{r}/\text{t}}^{H}\mathbf{\bar{F}}_{n,h}^{H}\right),
\end{align}
where the constant matrix $\mathbf{\bar{F}}_{n,v}=\mathbf{\bar{C}}_{\bm{\varsigma}_{[\imath]}[v]}\mathbf{A}_{n}-\mathbf{A}_{n}\mathbf{\bar{C}}_{\bm{\varsigma}_{[\imath]}[v]}\mathbf{A}_{n}$ with $\mathbf{A}_{n}=[\mathbf{I}_{n},\mathbf{0}_{N-n}]$. Also, the QoS constraint \eqref{30c} can be transformed into
\begin{align}
\label{32} P_{k}H_{[\imath],n}^{k,k}
\geq
\gamma_{\text{t}}\left(\sum\limits_{j\neq k,j\in\mathcal{K}_{\imath}}P_{j}
H_{[\imath],n}^{k,j}+\sigma^2\right),
\end{align}
where $H_{[\imath],n}^{k,j}=\sum_{n=1}^{N-1}p_{n}\text{Tr}
(\mathbf{A}_{n}\mathbf{H}_{j,\text{S}}\mathbf{V}_{[\imath],k}\mathbf{H}_{j,\text{S}}^{H}\mathbf{A}_{n}\mathbf{U}_{\text{r}/\text{t}})$. Note that when $p_{n}=1$ and $p_{m}=0$ ($m\neq n$) hold, the selection matrix $\mathbf{A}$ can be exactly determined, i.e., $\mathbf{A}=\mathbf{A}_{n}$. Thus, the problem (30) can be converted to the following AL form without selection matrices.
\begin{subequations}
\begin{align}
\label{33a} \min\limits_{\mathbf{E}_{[\imath]},\mathbf{P}_{[\imath]},\mathbf{U}_{\text{r}/\text{t}},\mathbf{V}_{[\imath],k},\mathbf{R}_{[\imath],\text{s}},p_{n},\rho_{2}}\quad &\text{Tr}(\mathbf{E}_{[\imath]}^{-1})+\frac{1}{2\rho_{2}}\sum_{n=1}^{N-1}( p_{n}-p_{n}^{2}) \\
\label{33b}\text{s.t.}\qquad &\eqref{16b},\eqref{16c},\eqref{18},\eqref{19d}-\eqref{19f},\eqref{30f},\eqref{32},\\
\label{33c}&\sum_{n=1}^{N-1}p_{n}=1.
\end{align}
\end{subequations}
Thereinto, when $\rho_{2}\rightarrow\infty$, the penalty term $p_{n}-p_{n}^{2}$ approaches $0$, which would be equivalent to integer constraint $p_{n}\in\{0,1\}$. In the inner loop, we adopt the AO framework to optimize the transmit power $\mathbf{P}_{[\imath]}$, the ISAC waveform $\mathbf{R}_{[\imath],\text{s}}$, the reflection/transmission coefficients $\mathbf{U}_{\text{r}/\text{t}}$ and the weight coefficient $p_{n}$.

\subsection{Joint Beamforming and Elements Optimization}
\subsubsection{Receive beamforming optimization}With fixed $\{\mathbf{P}_{[\imath]},\mathbf{U}_{\text{r}/\text{t}},\mathbf{R}_{[\imath],\text{s}},p_{n}\}$, problem (33) is equivalent to the following feasible detection problem with respect to $\{\mathbf{V}_{[\imath],k}\}$.
\begin{subequations}
\begin{align}
\label{34a} \mathop {{\rm{find}}}\quad &\mathbf{V}_{[\imath],k} \\
\label{34b}\text{s.t.} \quad &\text{rank}(\mathbf{V}_{[\imath],k})=1,\\
\label{34b} &\eqref{30f},\eqref{32},
\end{align}
\end{subequations}
which can be efficiently handled by using penalty-based rank-one relaxation method \cite{X.Yu_rank-one}. The converted problem is given by
\begin{subequations}
\begin{align}
\label{35a} \min\limits_{\mathbf{V}_{[\imath],k}}\quad &\frac{1}{2\rho_{1}}\left[\text{Tr}(\mathbf{V}_{[\imath],k})-
\|\mathbf{V}_{[\imath],k}^{[n-1]}\|_{2}-\text{Tr}(\mathbf{\bar{v}}_{[\imath],k}^{[n-1]}
(\mathbf{\bar{v}}_{[\imath],k}^{[n-1]})^{H}(\mathbf{V}_{[\imath],k}\mathbf{V}_{[\imath],k}^{[n-1]}))\right] \\
\label{35b}\text{s.t.} \quad &\eqref{30f},\eqref{32},
\end{align}
\end{subequations}
where $\mathbf{V}_{[\imath],k}^{[n-1]}$ is the given point determined by ($n-1$)-th iteration and $\mathbf{\bar{v}}_{[\imath],k}^{[n-1]}$ is the leading eigenvector of $\mathbf{V}_{[\imath],k}^{[n-1]}$. Note that the optimal $\mathbf{v}_{[\imath],k}$ can be obtained by carrying out SCA iterations with the accuracy $\epsilon_{v}$.


\subsubsection{Transmit power and sensing waveform optimization} With fixed $\{\mathbf{U}_{\text{r}/\text{t}},\mathbf{v}_{[\imath],k},p_{n}\}$, problem (33) is reduced to
\begin{subequations}
\begin{align}
\label{36a} \min\limits_{\mathbf{R}_{[\imath],\text{s}},\mathbf{E}_{[\imath]},\mathbf{P}_{[\imath]}}\quad &\text{Tr}(\mathbf{E}_{[\imath]}^{-1}) \\
\label{36b}\text{s.t.} \quad &\eqref{16b},\eqref{18},\eqref{20c},\eqref{32},
\end{align}
\end{subequations}
The optimal solutions $\{\mathbf{R}_{[\imath],\text{s}},\mathbf{P}_{[\imath]}\}$ can be easily obtained by solving the SDP problem (36).

\subsubsection{Reflection/transmission coefficient optimization} With fixed $\{\mathbf{R}_{[\imath],\text{s}},\mathbf{P}_{[\imath]},\mathbf{v}_{[\imath],k},p_{n}\}$ and the transformation of \eqref{22}, problem (33) can be re-expressed as
\begin{subequations}
\begin{align}
\label{37a} \min\limits_{\mathbf{E}_{[\imath]},\mathbf{U}_{\text{r}/\text{t}}}\quad &\text{Tr}(\mathbf{E}_{[\imath]}^{-1}) \\
\label{37b} \text{s.t.} \quad &\eqref{18},\eqref{19d}-\eqref{19f},\eqref{32},
\end{align}
\end{subequations}
which can be optimally solved by \textbf{Algorithm \ref{SCA-rank-one}}.

\subsubsection{Weight coefficient optimization}
With the fixed $\{\mathbf{R}_{[\imath],\text{s}},\mathbf{P}_{[\imath]},\mathbf{v}_{[\imath],k},\mathbf{U}_{\text{r}/\text{t}}\}$, the following AL problem is obtained.
\begin{subequations}
\begin{align}
\label{38a} \min\limits_{\mathbf{E}_{[\imath]},p_{n}}\quad &\text{Tr}(\mathbf{E}_{[\imath]}^{-1})
+\frac{1}{2\rho_{2}}\sum_{n=1}^{N-1}(p_{n}-p_{n}^{2}) \\
\label{38b} \text{s.t.} \quad &\eqref{18},\eqref{20c},\eqref{32},\eqref{33c}.
\end{align}
\end{subequations}
To handle the non-convex penalty term $p_{n}-p_{n}^{2}$, we adopt the first-order Taylor expansion to construct the linear upper bound approximation expression at $m$-th iteration, i.e.,
\begin{align}\label{39}
p_{n}-p_{n}^{2} \leq p_{n}-(p_{n}^{[m-1]})^{2}-2p_{n}^{[m-1]}(p_{n}-p_{n}^{[m-1]}),
\end{align}
where $p_{n}^{[m-1]}$ denotes the optimized value of $p_{n}$ at $(m-1)$-th iteration. By substituting the right-hand side of \eqref{39} into the penalty term of the objective function \eqref{38a}, the problem (38) becomes convex with respect to $p_{n}$ and can be solved over the SCA iterations. The details of the SCA algorithm to optimize weight coefficient is summarized in \textbf{Algorithm \ref{SCA-pn}}.

\begin{algorithm}[t]
    \caption{SCA algorithm for weight coefficient optimization.}
    \label{SCA-pn}
    \begin{algorithmic}[1]
        \STATE{Initialize $p_{n}^{[m-1]}$ and $\text{Tr}(\text{CRB}(\bm{\Psi}_{[\imath]}))^{[m-1]}$ with $m=1$. Set a convergence accuracy $\epsilon_{3}$.}
        \REPEAT
        \STATE{ update $p_{n}^{[m]}$ by solving problem (38).}
        \STATE{ calculate the $\text{Tr}(\text{CRB}(\bm{\Psi}_{[\imath]}))^{[m]}$ and set $m=m+1$.}
        \UNTIL{$|\text{Tr}(\text{CRB}(\bm{\Psi}_{[\imath]}))^{[m]}-\text{Tr}(\text{CRB}(\bm{\Psi}_{[\imath]}))^{[m-1]}|\leq \epsilon_{3}$.}
    \end{algorithmic}
\end{algorithm}

In the outer loop, we consider initializing $\rho_{2}$ as a large value to make the integer constraint trivial at the beginning, so that we can find a good starting point for original problem. Then, we gradually decrease $\rho_{2}$ over the outer loop iterations to obtain the feasible solution.

\subsection{Overall Algorithm}
The overall algorithm is summarized in \textbf{Algorithm \ref{PDL}}. Similarly, since the optimal transmit power, sensing waveform, reflection/transmission coefficients and weight coefficients are guaranteed at each step, the proposed PDL algorithm theoretically converges to the stationary point solution over the non-increasing iterations. For the computational complexity, the entire complexity of \textbf{Algorithm \ref{PDL}} relies on the complexity to solve the SDP problems (35), (36) (37) and (38). By exploiting the interior-point method, the computational complexity of PDL algorithm at the $\imath$-th phase can be expressed as $\mathcal{O}\Big(l_{\text{out}}\log(\frac{1}{\epsilon_{2}})\big(\log(\frac{1}{\epsilon_{v}})(M_\text{r}^{2})^{3.5}+(M_{\text{t}}^{2}+4+K_{\imath})^{3.5}+\log(\frac{1}{\epsilon_{1}})(N^{2}+4)^{3.5}
+ \log(\frac{1}{\epsilon_{3}})(N+3)^{3.5}\big)\Big)$ \cite{Z.Luo_complexity}, where $K_{\imath}=K_{1}$ in case of $\imath=1$, $K_{\imath}=K-K_{1}$ in case of $\imath=2$, and $l_{\text{out}}$ denotes the iteration number required in the outer loop.
\begin{algorithm}[t]
    \caption{Penalty-based double-loop algorithm.}
    \label{PDL}
    \begin{algorithmic}[1]
        \STATE{Initialize $\{\mathbf{U}_{\text{r}/\text{t}}^{[m-1]},p_{n}^{[m-1]},\text{Tr}(\text{CRB}(\bm{\Psi}_{[\imath]}))^{[m-1]}\}$ with $m=1$ and $\rho_{2}^{[l-1]}$ with $l=1$. Set the convergence accuracy $\{\epsilon_{2},\rho_{\text{th}}\}$.}
        \REPEAT
        \REPEAT
        \STATE{ update $\mathbf{v}_{[\imath],k}^{[m]}$ by solving problem (35).}
        \STATE{ update $\{\mathbf{R}_{[\imath],\text{s}}^{[m]},\mathbf{P}_{[\imath]}^{[m]}\}$ by solving problem (36).}
        \STATE{ update $\mathbf{U}_{\text{r}/\text{t}}^{[m]}$ by carrying out \textbf{Algorithm \ref{SCA-rank-one}} to solve problem (37).}
        \STATE{ update $p_{n}^{[m]}$ by carrying out \textbf{Algorithm \ref{SCA-pn}} to solve problem (38).}
        \STATE{ calculate the CRB value $\text{Tr}(\text{CRB}(\bm{\Psi}_{[\imath]}))^{[m]}$and set $m=m+1$.}
        \UNTIL{$|\text{Tr}(\text{CRB}(\bm{\Psi}_{[\imath]}))^{[m]}-\text{Tr}(\text{CRB}(\bm{\Psi}_{[\imath]}))^{[m-1]}|\leq \rho_{\text{th}}$.}
        \STATE{$\rho_{2}^{[l]}=\frac{\rho_{2}^{[l]}}{c_{2}}$ ($c_{2}>1$) and set $l = l+1$.}
        \UNTIL{The penalty term $\sum_{n=1}^{N-1}(p_{n}-p_{n}^{2})$ drops below $\epsilon_{2}$.}
    \end{algorithmic}
\end{algorithm}

\section{Numerical Results}\label{Section_5}

\begin{figure}[t]
  \centering
  \includegraphics[scale = 0.6]{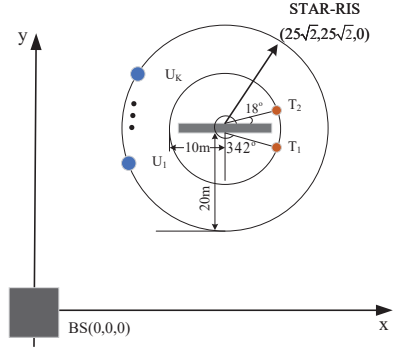}
  \caption{Top view of the simulation setup.}
  \label{Fig.2}
\end{figure}

In this section, the numerical results are provided to demonstrate the effectiveness of the proposed strategy. The top view of a three-dimensional coordinate network simulation setup is illustrated in Fig. \ref{Fig.2}, where the BS is located at $(0,0,0)$ m, and the STARS is located at a distance of $50$ m from the BS, i.e., $(25\sqrt{2},25\sqrt{2},0)$ m. $\text{U}_{k}$ is randomly distributed on a circle with a radius of $d_{k}$ m centred on STARS, while $\text{T}_{1}$ and $\text{T}_{1}$ are located at a distance of $10$ m from the STARS with the directions of $(342^{\circ},30^{\circ})$ and $(18^{\circ},30^{\circ})$. The path loss at the unit reference distance is set as $L_{0}=-30$ dB, the path-loss exponents for communication links are set as $2.2$, the path-loss exponents of sensing links are set as $2.5$, the noise power is set as $-115$ dBm, $T=10$, $N_{\text{v}}=5$, and $N_{\text{h}}=\frac{N}{N_{\text{v}}}$. The other simulation parameters are listed in the caption of each figure. Moreover, each figure is the average result over the $100$ independent Monte-Carlo experiments.

\subsection{Verification of Lemma \ref{Lemma_1} and Corollary \ref{Corollary_1}}
\begin{figure}[t]
\centering 
\begin{minipage}[t]{0.45\textwidth}
\includegraphics[width=1\textwidth]{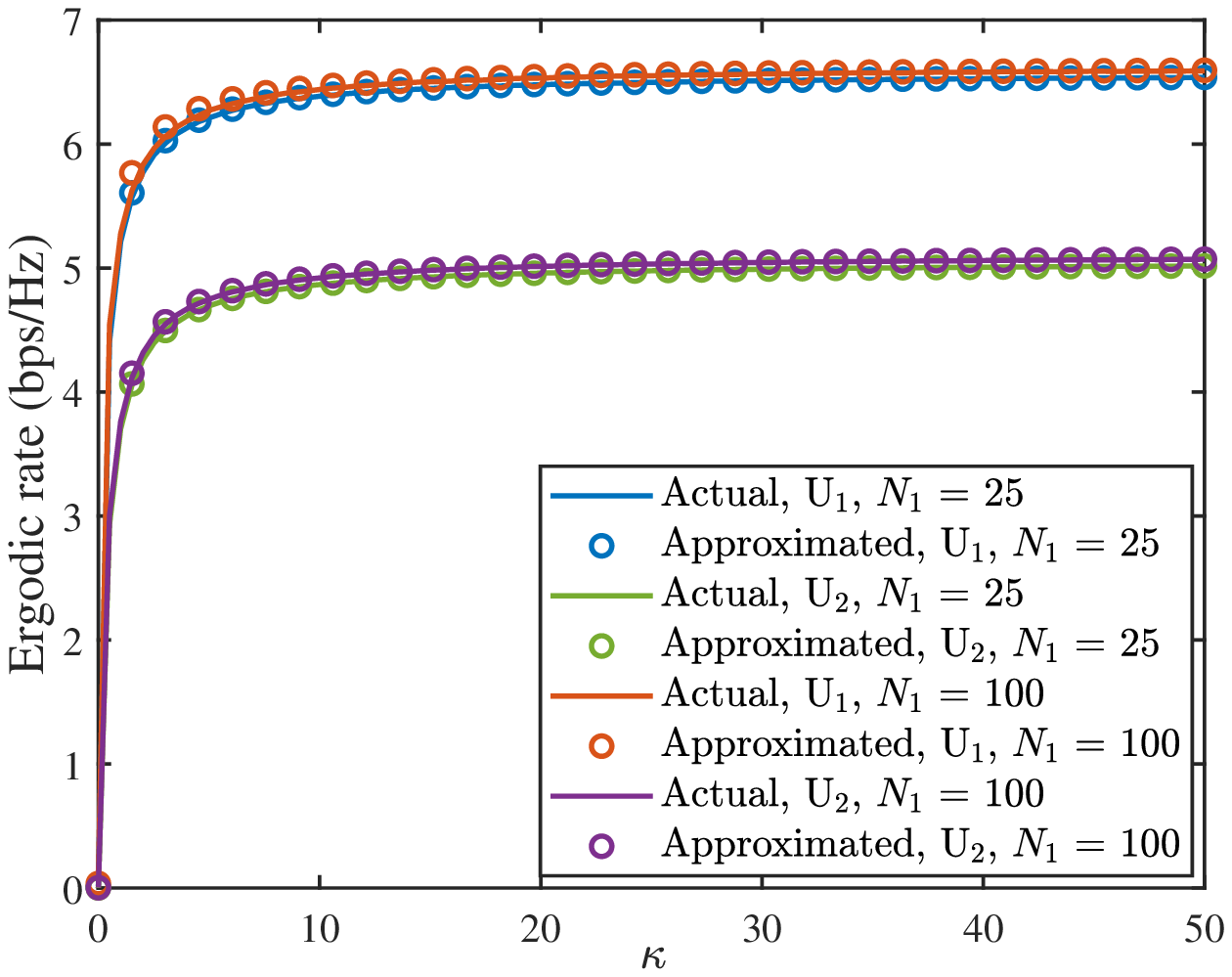}
\caption{Accuracy of the approximated ergodic rate versus $\kappa$ with $M_{\text{r}}=8$, $M_{\text{t}}=16$, $d_{1}=20$ m, $K=2$, $d_{2}=10$ m, and $P_{\text{U},\text{max}}=15$ dBm.}
\label{Fig.3}
\end{minipage}\qquad
\begin{minipage}[t]{0.45\textwidth} 
\centering 
\includegraphics[width=1\textwidth]{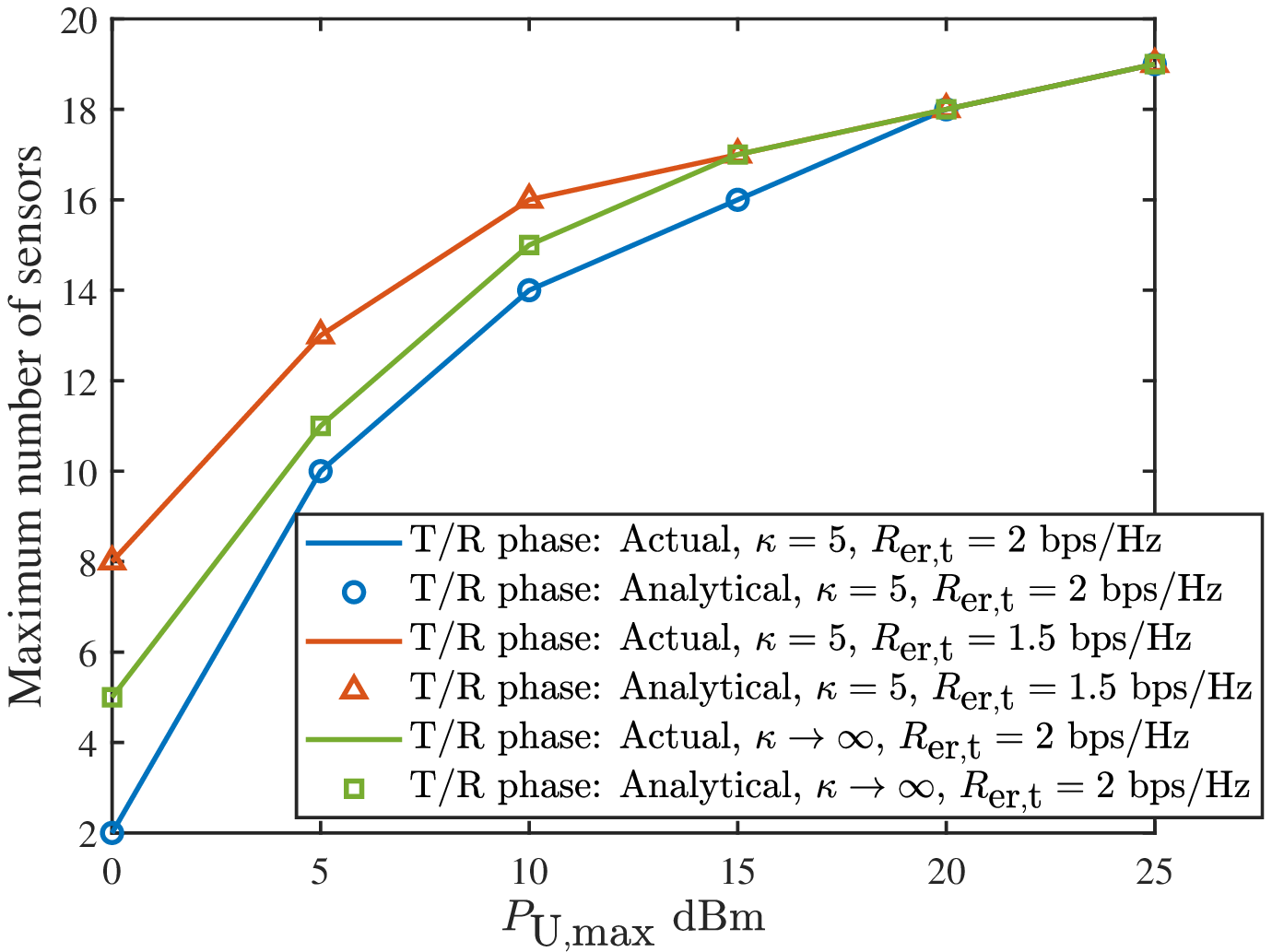} 
\caption{The maximum number of sensor deployment versus the transmit power budget at the users with $N=20$, $M_{\text{r}}=1$, $M_{\text{t}}=4$, $K=2$, $d_{k}=20$ m, and $P_{\text{BS},\text{max}}=15$ dBm.}
\label{Fig.4}
\end{minipage}
\end{figure}
The accuracy of the approximated ergodic rate expression derived in Lemma \ref{Lemma_1} is verified in Fig. \ref{Fig.3}, where the identity reflection/transmission coefficients $\mathbf{U}_{\text{r}/\text{t}}=\mathbf{I}_{N_{1}}$ are adopted and the numerical ergoric rate is calculated based on the $10000$ independent channels. It is shown that the approximated ergodic rate is accurate to the actual ergodic rate for different $N_{1}$, and $\text{U}_{2}$ achieves higher ergodic rate. These observations are expected since 1) the average non-LoS components follow the complex Gaussian distribution with unit variance due to the law of large numbers; and 2) $\text{U}_{1}$ suffers the severer large-scale path loss than $\text{U}_{2}$. It is also observed that the achievable ergodic rate firstly increases with the rise of $\kappa$ and gradually turns into stable. This is since that 1) when the $\kappa$ is relatively small, increasing $\kappa$ can make the LoS components prodominate and provide the stronger transmission links; and 2) when $\kappa$ becomes large, the Rician channels tend to be the constant pure LoS channels.

In Fig. \ref{Fig.4}, we evaluate the the maximum-number sensor deployment policy derived in Corollary \ref{Corollary_1} for both the transmission (T) and reflection (R) phases, where the actual maximum number of sensors are obtained by solving the feasible detection problem with different number of sensors for $100$ independent channel realizations. Firstly, we can observe that the analytical maximum number of sensors is exactly equal to the numerical results. Also can be seen, the maximum number of sensors for T phase and R phase are equal. This is intuitive because the maximum-number sensor deployment is only restricted by the corresponding QoS target rates, which are the same for different phases in the considered network. Moreover, when $P_{\text{U},\text{max}}$ increases and $R_{\text{er,\text{t}}}$ decreases, less PEs would be required to satisfy the QoS constraint, so a significant upward trend in the number of sensors can be observed.

\begin{figure}[t]
\centering 
\begin{minipage}[t]{0.45\textwidth}
\includegraphics[width=1\textwidth]{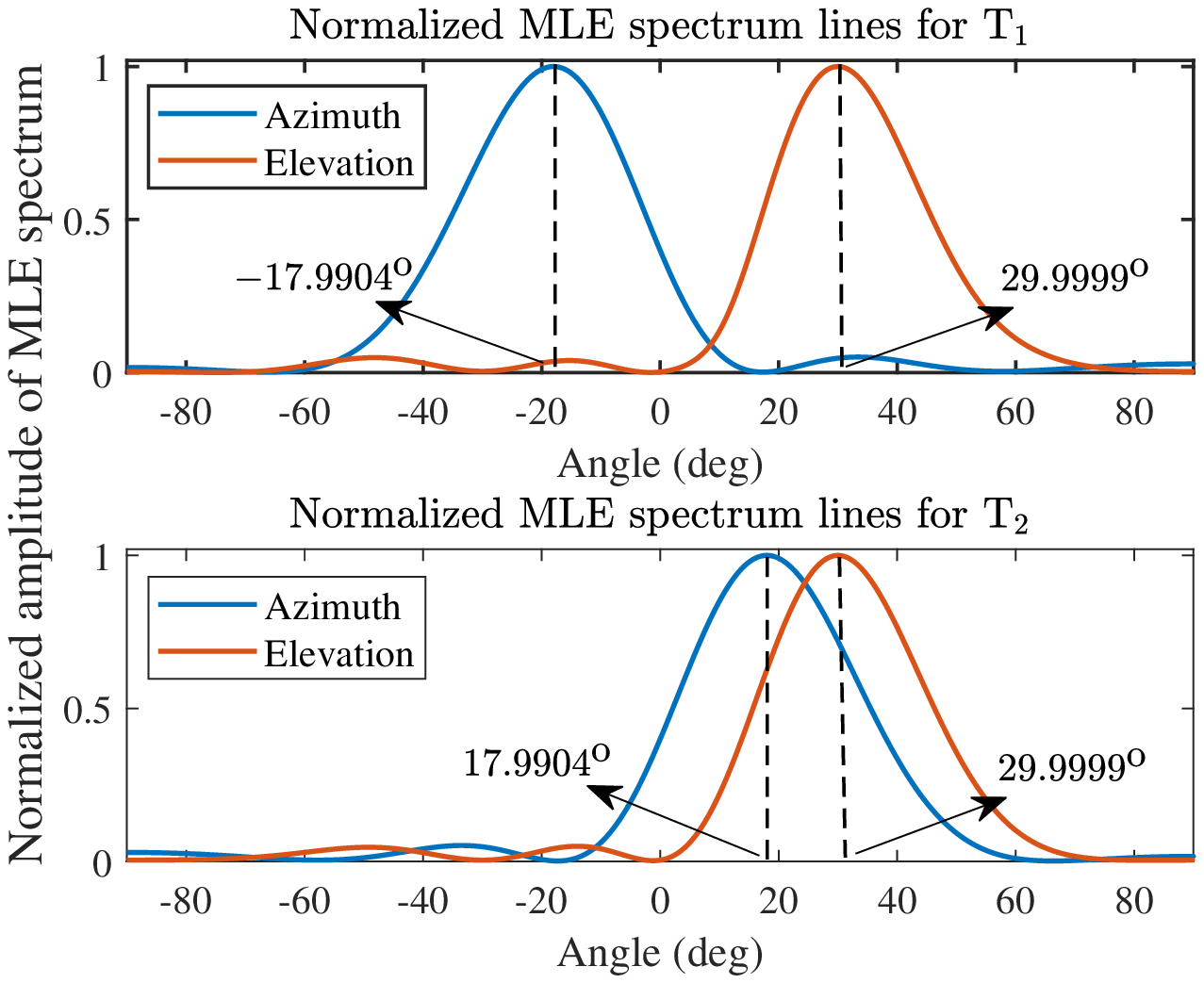}
\caption{The estimated azimuth and elevation angles via the MLE method with $N=20$, $N_{1}=5$, $M_{\text{r}}=M_{\text{t}}=8$, $K=2$, $d_{k}=20$ m, $R_{\text{er,\text{t}}}=2.5$ bps/Hz, and $P_{\text{U},\text{max}}=P_{\text{BS},\text{max}}=35$ dBm.}
\label{Fig.5}
\end{minipage}\qquad
\begin{minipage}[t]{0.45\textwidth} 
\centering 
\includegraphics[width=1\textwidth]{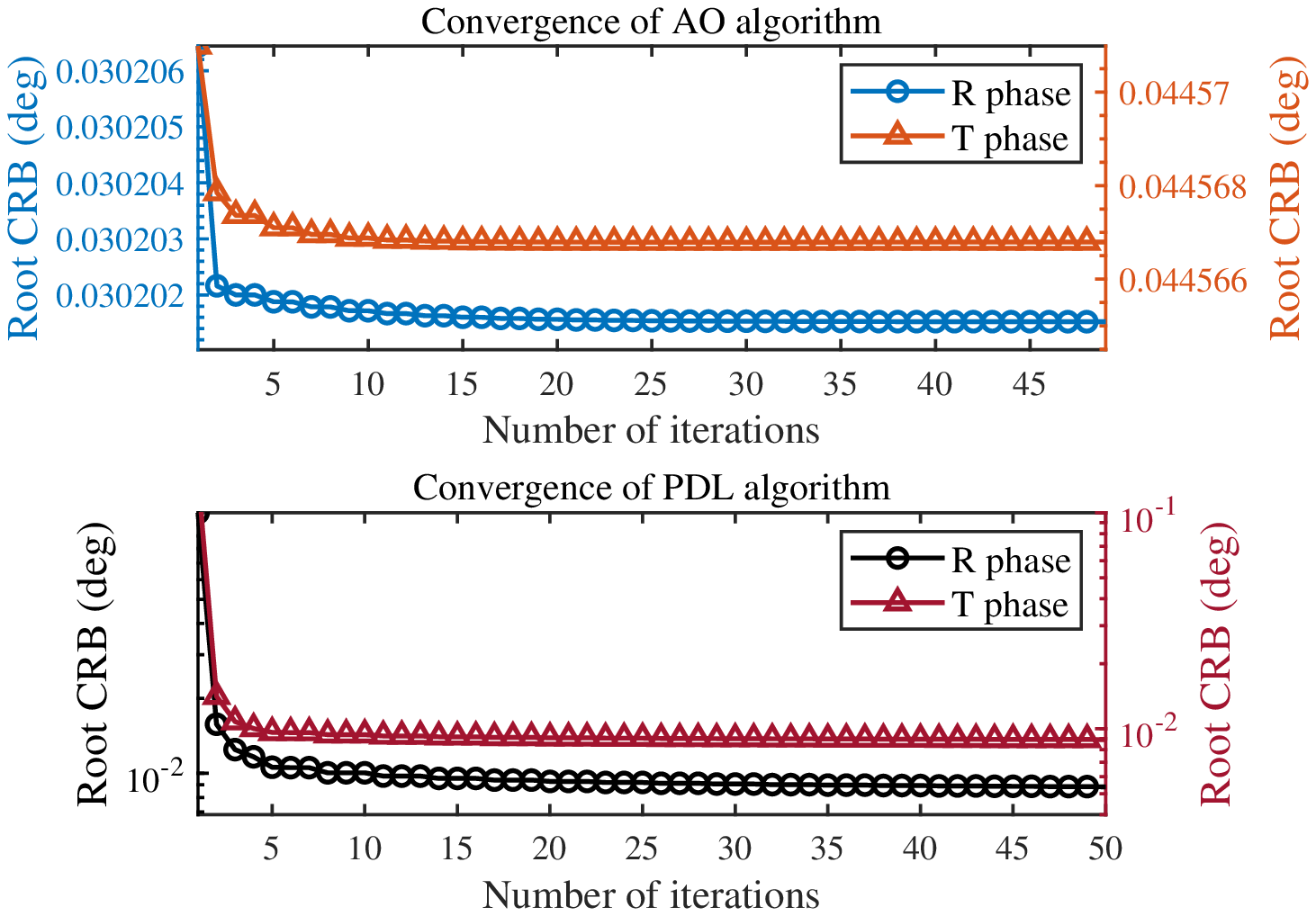} 
\caption{Convergence of the proposed algorithms with the common parameters $N=20$, $M_{\text{r}}=8$, and $d_{k}=20$ m.}
\label{Fig.6}
\end{minipage}
\end{figure}

\subsection{Effectiveness of Proposed Algorithms}
To demonstrate the correctness of the proposed algorithms from the perspective of sensing accuracy, the MLE spectrum lines under the two-user and fixed number of sensors are illustrated in Fig. \ref{Fig.5}, where the estimated DOA angle can be determined by the exhaustive search of the highest value point of the MLE spectrum \cite[Appendix E]{X.Song_IRS_ISAC}. As such, we can observe that the MLE estimates angles are $(-17.9904^{\circ},29.9999^{\circ})$ and $(17.9904^{\circ},29.9999^{\circ})$, respectively, which perfectly align with the presupposed DOA angles $(342^{\circ},30^{\circ})$ and $(18^{\circ},30^{\circ})$ and validate the effectiveness of the optimized sensing waveform and reflection/transmission coefficients. Fig. \ref{Fig.6} plots the convergence performance of the proposed algorithms, where $M_\text{t}=8$, $N_{1}=15$, $K=2$, $R_{\text{er,\text{t}}}=2.5$ bps/Hz, $P_{\text{U},\text{max}}=25$ dBm, $P_{\text{BS},\text{max}}=35$ dBm are adopted for AO algorithm, while $M_\text{t}=12$, $K=4$, $R_{\text{er,\text{t}}}=0.5$ bps/Hz, $P_{\text{U},\text{max}}=P_{\text{BS},\text{max}}=45$ dBm are set for PDL algorithm. Note that for coinciding with the practical DOA angles, we transform the radian-based root CRB into the degree unit in Fig. \ref{Fig.6}. As can be observed, both algorithms are capable of converging to the stationary point solutions within the finite iterations. This is because the optimal solutions at each step for solving the subproblem can be guaranteed in both two algorithms, which ensures a non-increasing trend over the iterations. Meanwhile, we can observe that R phase achieves smaller CRB compared to the T phase. It can be explained by the fact that in the R phase, the STARS reconfigure the communication signal from the users and the sensing signal from the BS as a new probing signal for estimation, which introduces more DoFs and enhance the received echo energy, thus improving the sensing performance.

\begin{figure}[t]
\centering 
\begin{minipage}[t]{0.45\textwidth} 
\centering 
\includegraphics[width=1\textwidth]{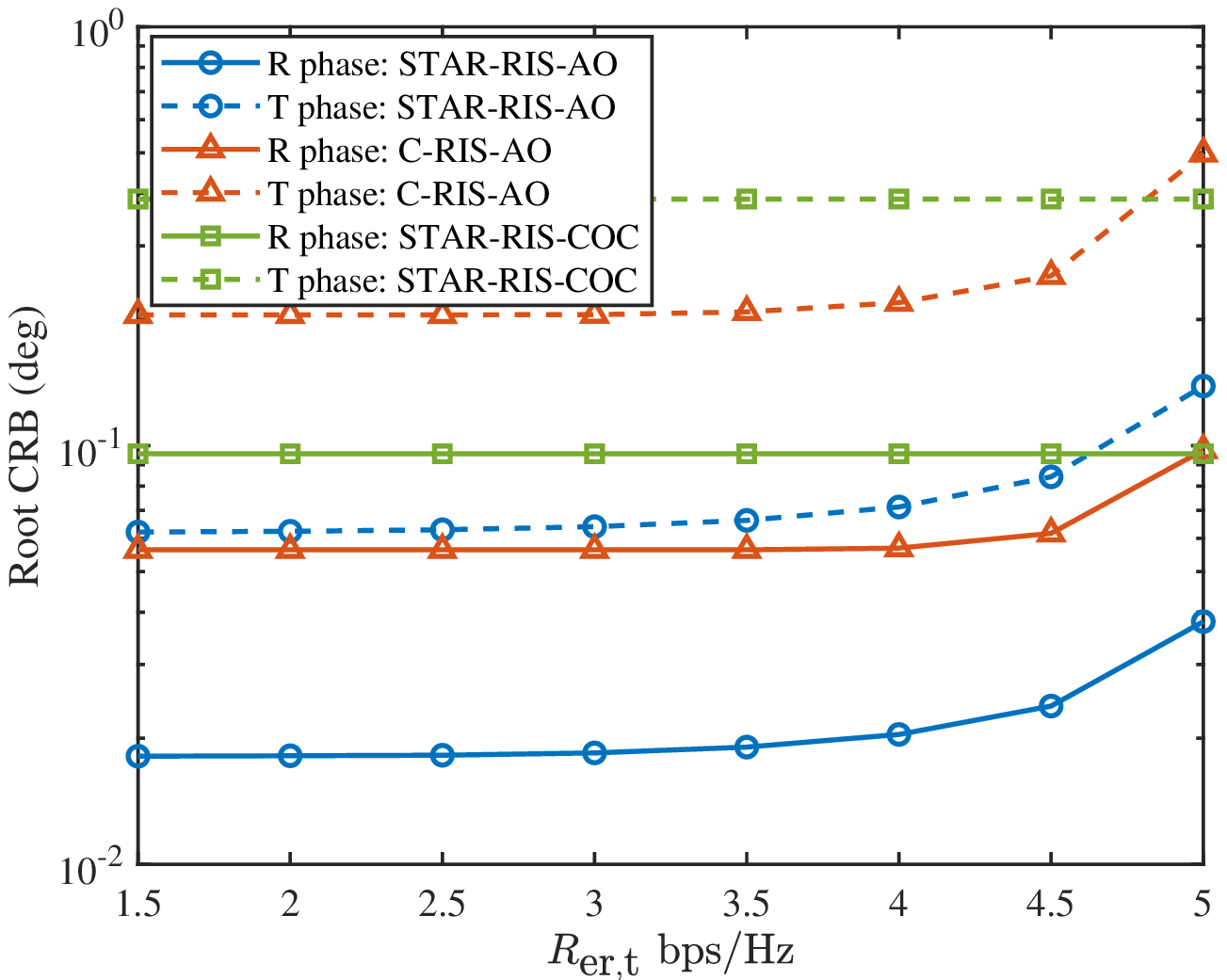}
\caption{The root CRB versus the QoS target rate with $N=20$, $N_{1}=10$, $M_{\text{r}}=M_{\text{t}}=8$, $K=2$, $d_{k}=20$ m, $P_{\text{U},\text{max}}=15$ dBm, and $P_{\text{BS},\text{max}}=30$ dBm .}
\label{Fig.7}
\end{minipage}\qquad
\begin{minipage}[t]{0.45\textwidth} 
\centering 
\includegraphics[width=1\textwidth]{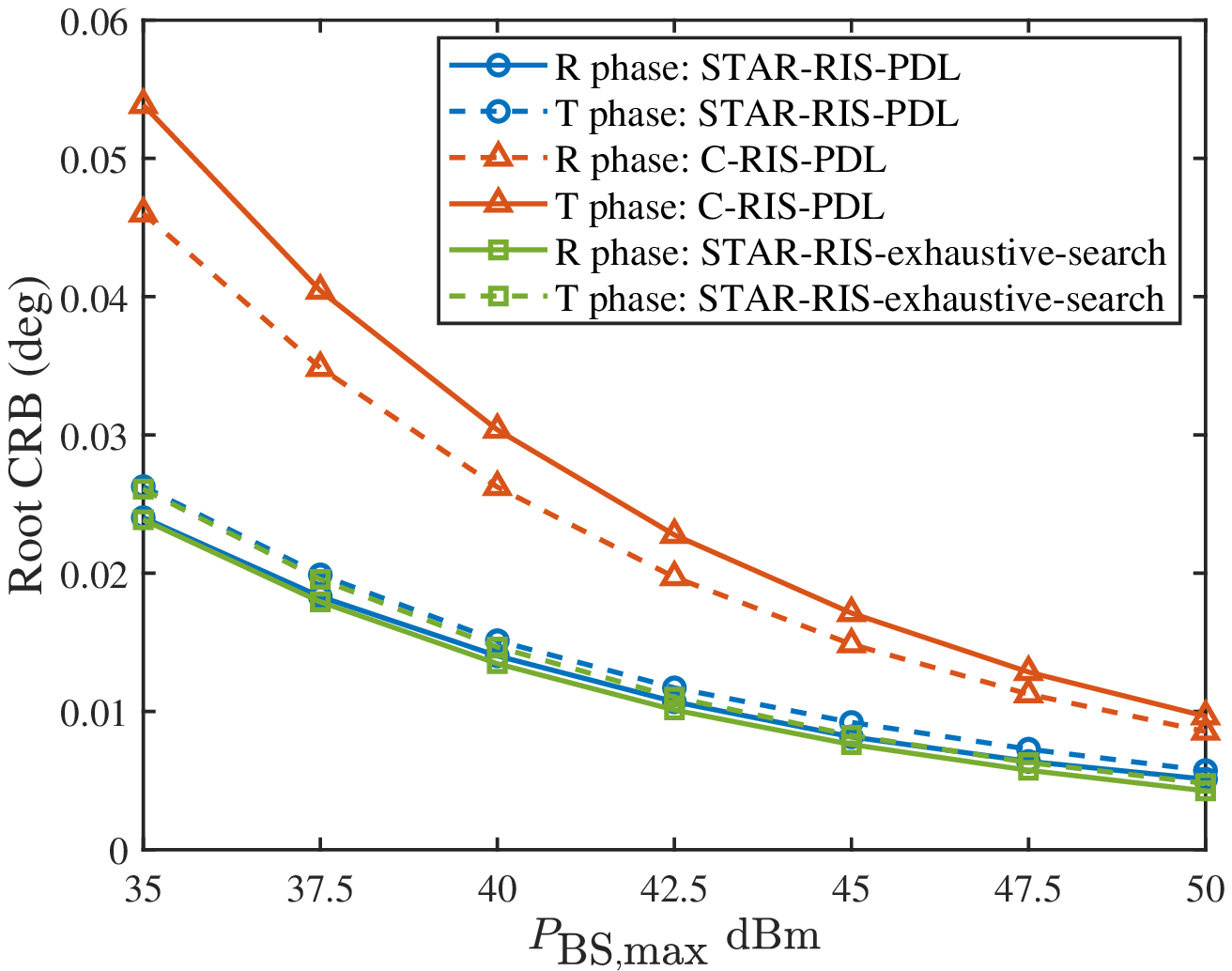} 
\caption{The root CRB versus the transmit power budget at the BS with with $N=20$, $M_{\text{r}}=8$, $M_{\text{r}}=12$, $K=4$, $d_{k}=20$ m, $R_{\text{er,\text{t}}}=1$ bps/Hz, and $P_{\text{U},\text{max}}=45$ dBm.}
\label{Fig.8}
\end{minipage}
\end{figure}

\subsection{Performance Comparison}
To verify the performance of the proposed scheme, we consider following benchmark schemes for comparison:
\begin{itemize}
  \item \textbf{C-RIS-AO}: In the ``conventional-RIS-AO''  (C-RIS-AO) scheme, we consider replacing the STARS by a reflecting-only RIS and a transmitting-only RIS, where the proposed AO algorithm is modified to jointly optimize the coefficients of PEs at conventional RIS, transmit power at users and the receive beamforming at the BS under the two-user and fixed sensor number setup. For fairness comparison, each conventional RIS is assumed to possess $\frac{N_{1}}{2}$ PEs and $\frac{N_{2}}{2}$ sensors.
  \item \textbf{STARS-COC}: In the ``STARS-communication-oriented coefficients (COC)'' scheme, the reflection/transmission coefficients of PEs is determined by maximizing the minimum ergodic rate of the user, i.e., $\bm{\Theta}_{\text{r}/\text{t}}^{\text{COC}}=\max\limits_{\bm{\Theta}_{\text{r}/\text{t}}}\ \min\limits_{k} \ \mathbb{E}\{R_{[\imath],k}\}$.
  \item \textbf{C-RIS-PDL}: In the ``C-RIS-PDL'' scheme, we deploy an $\frac{N}{2}$-element reflecting-only RIS and an $\frac{N}{2}$-element transmitting-only RIS at the same location of STARS, where the proposed PDL algorithm is modified to jointly optimize the number of sensors, the coefficients of PEs, transmit power at users and the receive beamforming at the BS.
  \item \textbf{STARS-exhaustive-search}: In the ``STARS-exhaustive-search'' scheme, we independently optimize $N-1$ subproblems with respect to the reflection/transmission coefficients, transmit power and the beamforming at the BS, where $p_{n}=1$ ($\sum_{n}^{N-1}p_{n}=1$) is set for the $n$-th subproblem. Then, we choose the smallest CRB as the final solution of this scheme.
\end{itemize}

As depicted in Fig. \ref{Fig.7} and Fig. \ref{Fig.8}, the proposed scheme can always achieve better sensing performance compared to the conventional RIS. This can be expected since the STARS enables the double number of elements, which introduces more spatial DoFs for supporting the communication and sensing. In Fig. \ref{Fig.7}, it is also observed that the sensing performance of the considered network deteriorates with the increasing QoS target rate, and the STARS-COC scheme achieves the worst performance. An intuitive explanation for the phenomenon is: 1) both the sensing and communication relies on the coefficients setup of PEs, so when the QoS rate increases, STARS has to prioritise support for communications to satisfy the QoS constraints, which leads to a reduction in sensing performance; and 2) since the STARS-COC scheme only focuses on the communication performance improvement, which would inevitably sacrifice the accuracy of sensing. Moreover, we can observe that the CRB achieved by the STARS-COC scheme does not vary with the QoS changes from Fig. \ref{Fig.7}. This is because the impact of QoS target rate on the sensing performance can only be realized by affecting the coefficients of PEs, which indicates that the sensing performance is independent of the QoS constraint under any fixed reflection/transmission coefficients setup. For the Fig. \ref{Fig.8}, we can observe that the proposed PDL algorithm can achieves the comparable performance of the optimal solution via the exhaustive search, which significantly validates the equivalence of the derived ECRB expression in Proposition 1.

\subsection{Impact of STARS Elements}

\begin{figure}[t]
\centering 
\begin{minipage}[t]{0.45\textwidth} 
\centering 
\includegraphics[width=1\textwidth]{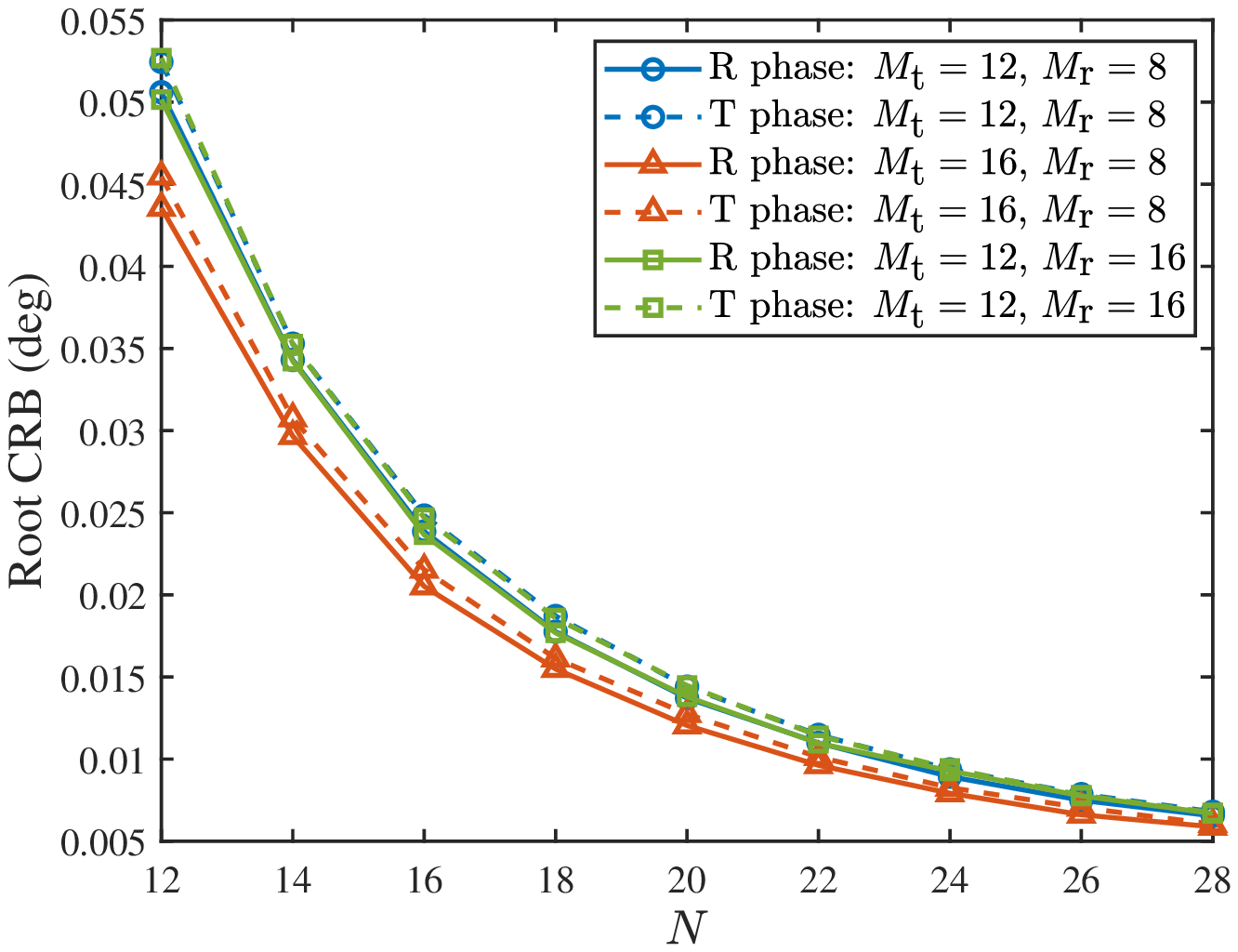}
\caption{The root CRB versus the number of STARS elements with $K=4$, $d_{k}=20$ m, $R_{\text{er,\text{t}}}=0.5$ bps/Hz, and $P_{\text{U},\text{max}}=P_{\text{BS},\text{max}}=45$ dBm.}
\label{Fig.9}
\end{minipage}\qquad
\begin{minipage}[t]{0.45\textwidth} 
\centering 
\includegraphics[width=1\textwidth]{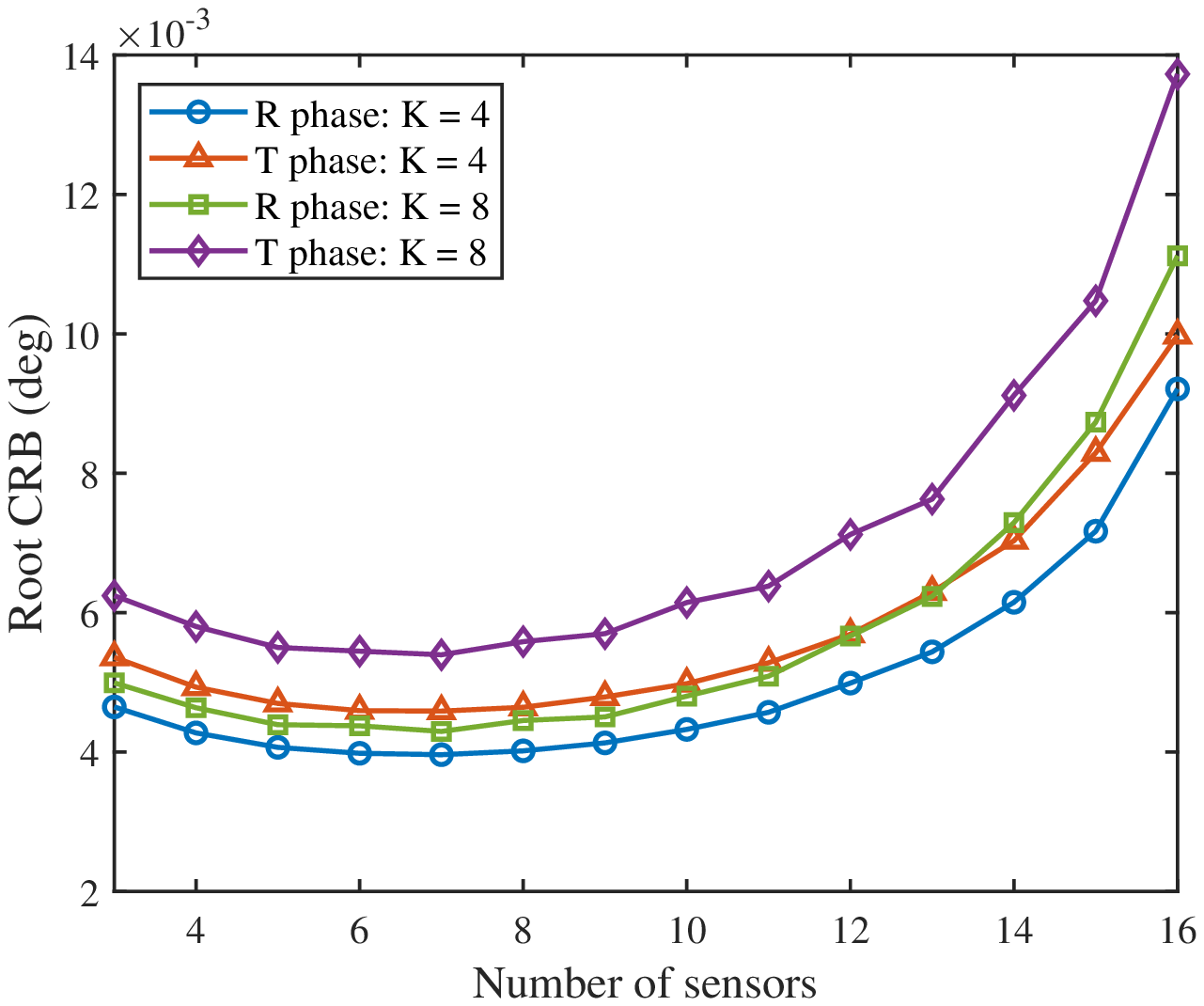} 
\caption{Benchmarks under 2-user setup with $N=20$, $M_{\text{r}}=8$, $M_{\text{t}}=12$, $d_{k}=20$ m, $R_{\text{er,\text{t}}}=0.5$ bps/Hz, $P_{\text{U},\text{max}}=P_{\text{BS},\text{max}}=45$ dBm, and $N_{\text{v}}=10$.}
\label{Fig.10}
\end{minipage}
\end{figure}

Fig. \ref{Fig.9} investigates the influence of STARS element number on the sensing performance. Firstly, It can be observed that the root CRB monotonically decreases as $N$ increases. This is because:1)a larger $N$ can provide a higher array gains; and 2) under the near-optimal allocation of number of sensors, increasing $N$ also increases the number of sensors, which improves the sensing reception ability at the sensing array. We can also observe an interesting result that increasing number of transmit antennas at the BS significantly reduces the root CRB, but increasing number of receive antennas at the BS has almost no effect on the CRB. This is resulted from the bi-directional sensing-STARS architecture. Specifically, by exploiting this architecture, the receive antenna at the BS is only responsible for receiving the communication signal, which implies that we only need the number of receive antennas that meets the QoS requirements in the considered network. However, increasing transmit antenna can expand the DoFs of the sensing waveform, which provides a more flexible design of the sensing waveform for supporting the sensing.

In Fig. \ref{Fig.10}, we illustrate the impact of the number of sensors under the fixed total number of STARS elements. It can be seen that when the number of sensors is less than 7, deploying more sensors than PEs are preferable, and when the number of sensors is larger than 7, increasing the number of sensors can hardly bring the constructive impact to the network anymore. It reveals that there exists a trade-off between the the number of sensors and PEs. In detail, with a small number of sensors, it is required to deploy more sensors to obtain sufficient echo sampling resolution to extract target information. Whereas, in the case of large number of sensors, the information loss of the sensing targets caused by insufficient echo sampling resolution is relatively small. Thus, deploying more PEs to provide more DoFs for both communication and sensing becomes the dominant factor in the sensing performance. Furthermore, we observe that increasing number of users degrades the sensing performance, which is expected since the reflection/transmission coefficients of PEs needs to be more oriented towards enhancing communications for satisfying the QoS requirements of users.

\section{Conclusion}\label{Section_6}
A new STARS-empowered ISAC scheme was proposed, where a bi-directional sensing-STARS architecture was devised to support the full-space communication and sensing tasks. Two efficient algorithms were developed to obtain the near-optimal solutions of the CRB minimization problems. The correctness and effectiveness of proposed scheme was demonstrated by the experiment results. It was also unveiled that: 1) STARS was capable of providing superior performance compared to the conventional RIS; 2) under the unique design of bi-directional sensing-STARS architecture, the number of receive antennas at the BS has little impact on the sensing performance; 3) increasing number of PEs is more appealing than sensors for sensing performance improvement. This work validated the potential of STARS in supporting full-space dual-functional transmissions, and revealed an endogenous tradeoff that how do we determine the number of sensors to be deployed. Both of them provided useful guidance for practical system design.

\section*{Appendix A: Derivation of FIM Matrix}
According \eqref{11}, the element matrix in FIM matrix $\mathbf{F}_{[i]}$ can be expressed as
\begin{align}
\label{A-1} \tag{A-1}
\mathbf{J}_{\bm{\Psi}_{[\imath]}\bm{\Psi}_{[\imath]}}=\frac{2}{\sigma^{2}}\begin{bmatrix}\Re\left(\frac{\partial\mathbf{q}_{[\imath]}^{H}}{\partial\varphi_{[\imath]}}
\frac{\partial\mathbf{q}_{[\imath]}}{\partial\varphi_{[\imath]}}\right) & \Re\left(\frac{\partial\mathbf{q}_{[\imath]}^{H}}{\partial\varphi_{[\imath]}}
\frac{\partial\mathbf{q}_{[\imath]}}{\partial\phi_{[\imath]}}\right)\\
\Re\left(\frac{\partial\mathbf{q}_{[\imath]}^{H}}{\partial\phi_{[\imath]}}
\frac{\partial\mathbf{q}_{[\imath]}}{\partial\varphi_{[\imath]}}\right) & \Re\left(\frac{\partial\mathbf{q}_{[\imath]}^{H}}{\partial\phi_{[\imath]}}
\frac{\partial\mathbf{q}_{[\imath]}}{\partial\phi_{[\imath]}}\right) \end{bmatrix},
\end{align}
where $\frac{\partial\mathbf{q}_{[\imath]}}{\partial \bm{\Psi}_{[\imath]}[h]}\!=\!\text{vec}\!\left(\!\alpha_{[\imath]}\left[\frac{\partial\mathbf{b}(\varphi_{[\imath]},\phi_{[\imath]})}{\partial \bm{\Psi}_{[\imath]}[h]}
\mathbf{a}^{T}
(\varphi_{[\imath]},\phi_{[\imath]})\!+\!\mathbf{b}(\varphi_{[\imath]},\phi_{[\imath]})\frac{\partial\mathbf{a}^{T}(\varphi_{[\imath]},\phi_{[\imath]})}{\partial \bm{\Psi}_{[\imath]}[h]}\right]\bm{\Theta}_{\text{r}/\text{t}}\mathbf{X}_{[\imath]}\right)$ ($1\leq h\leq 2$). With the derivative chain rule, we have
\begin{align}
\label{A-2} \tag{A-2}
\frac{\partial\bm{\epsilon}(\varphi_{[\imath]},\phi_{[\imath]})}{\partial \bm{\Psi}_{[\imath]}[h]}=\bm{\epsilon}(\varphi_{[\imath]},\phi_{[\imath]})\odot\bm{\dot{\epsilon}}_{\bm{\Psi}_{[\imath]}[h]},\quad \bm{\epsilon}(\varphi_{[\imath]},\phi_{[\imath]})\in\{\mathbf{a}(\varphi_{[\imath]},\phi_{[\imath]}),\mathbf{b}(\varphi_{[\imath]},\phi_{[\imath]})\},
\end{align}
where the $n$-th elements of $\bm{\dot{\epsilon}}_{\bm{\Psi}_{[\imath]}[h]}$ are given by
\begin{align}
\label{A-3} \tag{A-3}
\bm{\dot{\epsilon}}_{\bm{\Psi}_{[\imath]}[h]}[n]=\begin{cases}\jmath\bar{n}\pi\cos\phi_{[\imath]}\cos\varphi_{[\imath]}, \quad &\text{if}\  h=1,\\ \nonumber
\jmath(-\bar{n}\pi\sin\phi_{[\imath]}\sin\varphi_{[\imath]}+(n-1-N_{\text{h}}\bar{n})\pi\cos\phi_{[\imath]}), \quad &\text{if}\  h=2,
\end{cases}
\end{align}
Let $\mathbf{\dot{Q}}_{\bm{\Psi}_{[\imath]}[h]}=\frac{\partial\mathbf{b}(\varphi_{[\imath]},\phi_{[\imath]})}{\partial \bm{\Psi}_{[\imath]}[h]}
\mathbf{a}^{T}
(\varphi_{[\imath]},\phi_{[\imath]})\!+\!\mathbf{b}(\varphi_{[\imath]},\phi_{[\imath]})\frac{\partial\mathbf{a}^{T}(\varphi_{[\imath]},\phi_{[\imath]})}{\partial \bm{\Psi}_{[\imath]}[h]}$, we can rewrite $\mathbf{J}_{\bm{\Psi}_{[\imath]}\bm{\Psi}_{[\imath]}}$ as
\begin{align}
\label{A-4} \tag{A-4}
\mathbf{J}_{\bm{\Psi}_{[\imath]}\bm{\Psi}_{[\imath]}}=\frac{2T|\alpha_{[\imath]}|^{2}}{\sigma^{2}}\Re\left(\begin{bmatrix}
\text{Tr}(\mathbf{\dot{Q}}_{\varphi_{[\imath]}}\bm{\Theta}_{\text{r}/\text{t}}\mathbf{R}_{\mathbf{x}_{[\imath]}}
\bm{\Theta}_{\text{r}/\text{t}}^{H}\mathbf{\dot{Q}}_{\varphi_{[\imath]}}^{H}) &
\text{Tr}(\mathbf{\dot{Q}}_{\varphi_{[\imath]}}\bm{\Theta}_{\text{r}/\text{t}}\mathbf{R}_{\mathbf{x}_{[\imath]}}
\bm{\Theta}_{\text{r}/\text{t}}^{H}\mathbf{\dot{Q}}_{\phi_{[\imath]}}^{H}) \\
\text{Tr}(\mathbf{\dot{Q}}_{\phi_{[\imath]}}\bm{\Theta}_{\text{r}/\text{t}}\mathbf{R}_{\mathbf{x}_{[\imath]}}
\bm{\Theta}_{\text{r}/\text{t}}^{H}\mathbf{\dot{Q}}_{\varphi_{[\imath]}}^{H}) & \text{Tr}(\mathbf{\dot{Q}}_{\phi_{[\imath]}}\bm{\Theta}_{\text{r}/\text{t}}\mathbf{R}_{\mathbf{x}_{[\imath]}}
\bm{\Theta}_{\text{r}/\text{t}}^{H}\mathbf{\dot{Q}}_{\phi_{[\imath]}}^{H})\end{bmatrix}\right),
\end{align}
where $\mathbf{R}_{\mathbf{X}_{[\imath]}}\approx\frac{1}{T}\mathbf{X}_{[\imath]}\mathbf{X}_{[\imath]}^{H}$. Similarly, let $\mathbf{Q}_{[\imath]}=\mathbf{b}(\varphi_{[\imath]},\phi_{[\imath]})\mathbf{a}^{T}
(\varphi_{[\imath]},\phi_{[\imath]})$, we can obtain
\begin{align}
\label{A-5} \tag{A-5}
\mathbf{J}_{\bm{\Psi}_{[\imath]}\bm{\alpha}_{[\imath]}}=\frac{2T}{\sigma^{2}}\Re\left(\begin{bmatrix}
\tilde{\alpha}_{[\imath]}\text{Tr}(\mathbf{Q}_{[\imath]}\bm{\Theta}_{\text{r}/\text{t}}\mathbf{R}_{\mathbf{x}_{[\imath]}}
\bm{\Theta}_{\text{r}/\text{t}}^{H}\mathbf{\dot{Q}}_{\varphi_{[\imath]}}^{H}) &
\jmath\tilde{\alpha}_{[\imath]}\text{Tr}(\mathbf{Q}_{[\imath]}\bm{\Theta}_{\text{r}/\text{t}}\mathbf{R}_{\mathbf{x}_{[\imath]}}
\bm{\Theta}_{\text{r}/\text{t}}^{H}\mathbf{\dot{Q}}_{\varphi_{[\imath]}}^{H}) \\
\tilde{\alpha}_{[\imath]}\text{Tr}(\mathbf{Q}_{[\imath]}\bm{\Theta}_{\text{r}/\text{t}}\mathbf{R}_{\mathbf{x}_{[\imath]}}
\bm{\Theta}_{\text{r}/\text{t}}^{H}\mathbf{\dot{Q}}_{\varphi_{[\imath]}}^{H}) & \jmath\tilde{\alpha}_{[\imath]}\text{Tr}(\mathbf{Q}_{[\imath]}\bm{\Theta}_{\text{r}/\text{t}}\mathbf{R}_{\mathbf{x}_{[\imath]}}
\bm{\Theta}_{\text{r}/\text{t}}^{H}\mathbf{\dot{Q}}_{\varphi_{[\imath]}}^{H}) \end{bmatrix}\right),
\end{align}
\begin{align}
\label{A-6} \tag{A-6}
\mathbf{J}_{\bm{\alpha}_{[\imath]}\bm{\alpha}_{[\imath]}}=\frac{2T}{\sigma^{2}}
\text{Tr}(\mathbf{Q}_{[\imath]}\bm{\Theta}_{\text{r}/\text{t}}\mathbf{R}_{\mathbf{x}_{[\imath]}}
\bm{\Theta}_{\text{r}/\text{t}}^{H}\mathbf{Q}_{[\imath]}^{H})\mathbf{I}_{2}.
\end{align}
This completes the derivation.

\section*{Appendix B: Derivation of Ergodic Rate}
With the results in \cite[Lemma 1]{Q.Zhang_ergodic}, the ergodic rate is approximated as
\begin{align}
\label{B-1} \tag{B-1}
\mathbb{E}\{R_{[\imath],k}\}\approx\frac{1}{2}\log_{2}\left(1+\frac{P_{k}\mathbb{E}\{\|\mathbf{h}_{k,\text{S}}^{H}\bm{\Theta}_{\text{r}/\text{t}}\mathbf{G}_{\text{r}}\|^2\}}
{\sigma^2}\right).
\end{align}
Substituting \eqref{1} into \eqref{B-1}, we can obatin
\begin{align}
 \nonumber
\mathbb{E}\{\|\mathbf{h}_{k,\text{S}}^{H}\bm{\Theta}_{\text{r}/\text{t}}\mathbf{G}_{\text{r}}\|^2\}\!&=\!
\mathbb{E}\!\left\{\left\|\Bigg(\sqrt{\frac{\kappa L_{k,\text{S}}^{2}}{1+\kappa}}\mathbf{\hat{h}}_{k,\text{S}}^{H}\!+\!\sqrt{\frac{L_{k,\text{S}}^{2}}{1+\kappa}}\mathbf{\tilde{h}}_{k,\text{S}}^{H}\Bigg)
\bm{\Theta}_{\text{r}/\text{t}}\Bigg(\sqrt{\frac{\kappa L_{\text{r}}^{2}}{1+\kappa}}\mathbf{\hat{G}}_{\text{r}}\!+\!\sqrt{\frac{L_{\text{r}}^{2}}{1+\kappa}}\mathbf{\tilde{G}}_{\text{r}}\Bigg)\right\|^2\right\},\\  \nonumber
&\overset{(a)}{=} \frac{L_{k,\text{S}}^{2}L_{\text{r}}^{2}}{(1+\kappa)^{2}}\Big(\kappa^{2}\mathbb{E}\{\|\mathbf{\hat{h}}_{k,\text{S}}^{H}\bm{\Theta}_{\text{r}/\text{t}}\mathbf{\hat{G}}_{\text{r}}\|^{2}\}+
\kappa \mathbb{E}\{\|\mathbf{\hat{h}}_{k,\text{S}}^{H}\bm{\Theta}_{\text{r}/\text{t}}\mathbf{\tilde{G}}_{\text{r}}\|^{2}\}+ \\ \nonumber
&\qquad\qquad\qquad\kappa \mathbb{E}\{\|\mathbf{\tilde{h}}_{k,\text{S}}^{H}\bm{\Theta}_{\text{r}/\text{t}}\mathbf{\hat{G}}_{\text{r}}\|^{2}\}+
\mathbb{E}\{\|\mathbf{\tilde{h}}_{k,\text{S}}^{H}\bm{\Theta}_{\text{r}/\text{t}}\mathbf{\tilde{G}}_{\text{r}}\|^{2}\}\Big), \\ \label{B-2}\tag{B-2}
&\overset{(b)}{=} \frac{\Big(\kappa^{2}\|\mathbf{\hat{h}}_{k,\text{S}}^{H}\bm{\Theta}_{\text{r}/\text{t}}\mathbf{\hat{G}}_{\text{r}}\|^{2}\!+\!
\kappa M_{\text{r}}\|\mathbf{\hat{h}}_{k,\text{S}}^{H}\bm{\Theta}_{\text{r}/\text{t}}\|^{2}\!+\!\kappa  \|\bm{\Theta}_{\text{r}/\text{t}}\mathbf{\hat{G}}_{\text{r}}\|^{2}_{\text{F}}\!+\!
M_{\text{r}}\|\bm{\Theta}_{\text{r}/\text{t}}\|^{2}_{\text{F}}\Big)}{\sfrac{(1+\kappa)^{2}}{L_{k,\text{S}}^{2}L_{\text{r}}^{2}}},
\end{align}
where equality (a) holds because the $\mathbf{\tilde{h}}_{k,\text{S}}$ and $\mathbf{\tilde{G}}_{\text{r}}$ are independent of each other, while equality (b) holds since all the elements in $\mathbf{\tilde{h}}_{k,\text{S}}$ and $\mathbf{\tilde{G}}_{\text{r}}$ follow the complex Gaussian distribution with zero mean and unit variance. With the result of \eqref{B-2}, we can easily obtain the approximated ergodic rate expression in Lemma \ref{Lemma_1}. This completes the proof.

\section*{Appendix C: Proof of Corollary \ref{Corollary_1}}
For the considered problem (16), we can readily know that achieving the maximum-number sensor deployment is tantamount to search for the minimum-dimensional reflection/transmission coefficients and the feasible transmit power that satisfy the QoS constraint, i.e.,
    \begin{subequations}
    \begin{align}
    \label{C-1a}\tag{C-1a} &\mathop {{\rm{find}}}  \quad  \{\bm{\Theta}_{\text{r}/\text{t}}^{\text{min}},\mathbf{P}\}\\
    \label{C-2b}\tag{C-2b} &\quad\text{s.t.} \quad \eqref{16d},\eqref{16f},\eqref{16g},
    \end{align}
    \end{subequations}
    where $\bm{\Theta}_{\text{r}/\text{t}}^{\text{min}}$ denotes the minimum-dimensional coefficient matrix. By substituting $M_{\text{r}}=1$ to the ergodic rate expression in Lemma \ref{Lemma_1}, it is readily to show that when the reflection/transmission coefficients of PEs are aligned to the cascaded channels $\mathbf{\hat{h}}_{k,\text{S}}^{H}\bm{\Theta}_{\text{r}/\text{t}}\mathbf{\hat{g}}_{\text{r}}$, i.e., the reflection/transmission coefficients derived in \eqref{24}, and $P_{k} = P_{\text{U},\text{max}}$ holds, it achieves the best communication performance with the least $N_{1}$. As such, constraint \eqref{19c} can be transformed into
    \begin{equation}\label{C-3}\tag{C-3}
    \frac{\kappa^{2}}{(1+\kappa)^{2}}N_{1}^{2}+
    \frac{2\kappa+1}{(1+\kappa)^{2}}N_{1} -\frac{(2^{2R_{\text{er},\text{t}}}-1)\sigma^{2}}{P_{\text{U},\text{max}}L_{k,\text{S}}^{2}L_{\text{r}}^{2}}\geq 0.
    \end{equation}
    Resorting the standard quadratic-root formula and the non-negativity of $N_{1}$, the minimum $N_{1}^{\text{min}}$ can be determined. Thus, the maximum $N_{2}^{\text{max}}$ can be derived as shown in \eqref{25} based on $N_{1}^{\text{min}}+N_{2}^{\text{max}}=N$. This completes the proof.

\section*{Appendix D: Derivation of Extended FIM Matrix}
Firstly, we can rewrite \eqref{11} as
\begin{align}
\label{D-1} \tag{D-1}\mathbf{y}_{[\imath],\text{s}}=\underbrace{\text{vec}\big(\alpha_{[\imath]}\mathbf{B}\bm{\varepsilon}_{[\imath]}\bm{\varepsilon}_{[\imath]}^{T}\mathbf{A}
\bm{\Theta}_{\text{r}/\text{t}}\mathbf{X}_{[\imath]}}_{\mathbf{q}_{[\imath]}}\big)+\mathbf{n}_{[\imath]},
\end{align}
where $\mathbf{X}_{[\imath]}\in\mathbb{C}^{N\times T}$ is the equivalent signal matrix with regarding the whole STARS as PEs. Similar to Appendix A, we have
\begin{align}
\label{D-2} \tag{D-2}
\frac{\partial\mathbf{q}_{[\imath]}}{\partial \bm{\Psi}_{[\imath]}[h]}=\text{vec}\left(\alpha_{[\imath]}\mathbf{B}\left[\frac{\partial\bm{\varepsilon}_{[\imath]}}{\partial \bm{\Psi}_{[\imath]}[h]}\bm{\varepsilon}_{[\imath]}^{T}+\bm{\varepsilon}_{[\imath]}
\frac{\partial\bm{\varepsilon}_{[\imath]}^{T}}{\partial \bm{\Psi}_{[\imath]}[h]}\right]\mathbf{A}\bm{\Theta}_{\text{r}/\text{t}}\mathbf{X}_{[\imath]}\right),\quad 1\leq h\leq 2,
\end{align}
where $\frac{\partial\bm{\varepsilon}_{[\imath]}}{\partial \bm{\Psi}_{[\imath]}[h]}$ is given in \eqref{A-2} and \eqref{A-3} with $1\leq n\leq N$. Hence, the $h$-th row and $v$-th column element of $\mathbf{J}_{\bm{\Psi}_{[\imath]}\bm{\Psi}_{[\imath]}}$ is given by
\begin{align}
\nonumber
\frac{\partial\mathbf{q}_{[\imath]}^{H}}{\partial \bm{\Psi}_{[\imath]}[h]}\frac{\partial\mathbf{q}_{[\imath]}}{\partial \bm{\Psi}_{[\imath]}[v]}&=\text{vec}\big(\alpha_{[\imath]}\mathbf{B}\mathbf{\dot{C}}_{\bm{\Psi}_{[\imath]}[h]}\mathbf{A}\bm{\Theta}_{\text{r}/\text{t}}\mathbf{X}_{[\imath]}\big)^{H}
\text{vec}\big(\alpha_{[\imath]}\mathbf{B}\mathbf{\dot{C}}_{\bm{\Psi}_{[\imath]}[v]}\mathbf{A}\bm{\Theta}_{\text{r}/\text{t}}\mathbf{X}_{[\imath]}\big),\\ \label{D-3} \tag{D-3}
&=T|\alpha_{[\imath]}|^{2}\text{Tr}\left(\mathbf{B}\mathbf{\dot{C}}_{\bm{\Psi}_{[\imath]}[v]}\mathbf{A}\bm{\Theta}^{\text{r}/\text{t}}\mathbf{R}_{\mathbf{X}_{[\imath]}}
    \bm{\Theta}_{\text{r}/\text{t}}^{H}\mathbf{A}\mathbf{\dot{C}}_{\bm{\Psi}_{[\imath]}[h]}^{H}\mathbf{B}\right), \ 1\leq h,v\leq 2,
\end{align}
where $\mathbf{\dot{C}}_{\bm{\Psi}_{[\imath]}[h]}=\frac{\partial\bm{\varepsilon}_{[\imath]}}{\partial \bm{\Psi}_{[\imath]}[h]}\bm{\varepsilon}_{[\imath]}^{T}+\bm{\varepsilon}_{[\imath]}
\frac{\partial\bm{\varepsilon}_{[\imath]}^{T}}{\partial \bm{\Psi}_{[\imath]}[h]}$. In the same way, the $h$-th row and $v$-th column element of $\mathbf{J}_{\bm{\Psi}_{[\imath]}\bm{\alpha}_{[\imath]}}$ and the $h$-th element on the diagonal of $\mathbf{J}_{\bm{\alpha}_{[\imath]}\bm{\alpha}_{[\imath]}}$ can be rewritten as
\begin{align}
\label{D-4} \tag{D-4}
\frac{\partial\mathbf{q}_{[\imath]}^{H}}{\partial \bm{\Psi}_{[\imath]}[h]}\frac{\partial\mathbf{q}_{[\imath]}}{\partial \bm{\alpha}_{[\imath]}[v]}&=T(\jmath)^{v-1}\tilde{\alpha}_{[\imath]}\text{Tr}\left(\mathbf{B}\mathbf{C}_{[\imath]}\mathbf{A}\bm{\Theta}^{\text{r}/\text{t}}\mathbf{R}_{\mathbf{X}_{[\imath]}}
    \bm{\Theta}_{\text{r}/\text{t}}^{H}\mathbf{A}\mathbf{\dot{C}}_{\bm{\Psi}_{[\imath]}[h]}^{H}\mathbf{B}\right), 1\leq h,v\leq 2,
\end{align}
\begin{align}
\label{D-5} \tag{D-5}
\frac{\partial\mathbf{q}_{[\imath]}^{H}}{\partial \bm{\alpha}_{[\imath]}[h]}\frac{\partial\mathbf{q}_{[\imath]}}{\partial \bm{\alpha}_{[\imath]}[h]}&=T\text{Tr}\left(\mathbf{B}\mathbf{C}_{[\imath]}\mathbf{A}\bm{\Theta}^{\text{r}/\text{t}}\mathbf{R}_{\mathbf{X}_{[\imath]}}
    \bm{\Theta}_{\text{r}/\text{t}}^{H}\mathbf{A}\mathbf{C}_{[\imath]}^{H}\mathbf{B}\right), 1\leq h\leq 2,
\end{align}
where $\mathbf{C}_{[\imath]}=\bm{\varepsilon}_{[\imath]}\bm{\varepsilon}_{[\imath]}^{T}$. Substituting \eqref{D-3}--\eqref{D-5} into \eqref{A-4}--\eqref{A-6}, we can obtain the extended FIM matrix in Proposition 1. Then, we prove the equivalence between the extended FIM matrix $\mathbf{F}_{[\imath]}$ and the original FIM matrix $\mathbf{F}_{[\imath]}^{\text{o}}$ under any given $\mathbf{A}$ and $\mathbf{B}$. To elaborate, let
$\mathbf{b}^{\text{p}}\in\mathbb{C}^{N_{2}\times 1}$ and $\mathbf{a}^{\text{p}}\in\mathbb{C}^{N_{1}\times 1}$ denote the practical steering vectors at the sensors and PEs, the received echo signal can be expressed as $\mathbf{E}_{[\imath]}^{\text{p}}=\alpha_{[\imath]}\mathbf{b}^{\text{p}}(\mathbf{a}^{\text{p}})^{T}\bm{\Theta}_{\text{r}/\text{t}}\mathbf{X}_{[\imath]}$. With this in hand, it is easy to obtain the following identity.
\begin{align}
\label{D-6} \tag{D-6}
\left\|\frac{\partial\mathbf{q}_{[\imath]}}{\bm{\varsigma}_{[\imath]}[i]}\right\|=\left\|\frac{\partial\text{vec}\left(\left[\mathbf{0}, \mathbf{0};
\mathbf{E}_{[\imath]}^{\text{p}}, \mathbf{0}\right]\right)}{\bm{\varsigma}_{[\imath]}[i]}\right\|
=\left\|\frac{\partial\mathbf{q}_{[\imath]}^{\text{p}}}{\bm{\varsigma}_{[\imath]}[i]}\right\|, \quad 1\leq i\leq 4,
\end{align}
where $\mathbf{q}_{[\imath]}^{\text{p}}=\text{vec}(\mathbf{E}_{[\imath]}^{\text{p}})$. Therefore, the $h$-th row and $v$-th column element of the extended FIM matrix is exactly equivalent to that of the original FIM matrix, i.e.,
\begin{align}
\label{D-7} \tag{D-7}
\mathbf{F}_{[\imath]}[h,v]=\frac{\partial\mathbf{q}_{[\imath]}^{H}}{\bm{\varsigma}_{[\imath]}[h]}\frac{\partial\mathbf{q}_{[\imath]}}{\partial \bm{\varsigma}_{[\imath]}[v]}=
\frac{\partial(\mathbf{q}_{[\imath]}^{\text{p}})^{H}}{\bm{\varsigma}_{[\imath]}[h]}\frac{\partial\mathbf{q}_{[\imath]}^{\text{p}}}{\partial \bm{\varsigma}_{[\imath]}[v]}=\mathbf{F}_{[\imath]}^{\text{o}}[h,v], \quad 1\leq h,v\leq 4.
\end{align}
This completes the proof.

\end{document}